\documentclass[twocolumn,pra,superscriptaddress,nofootinbib]{revtex4}
\pdfoutput=1
\usepackage{amsmath}
\usepackage{amssymb}
\usepackage{graphicx}
\newcommand{\ket}[1]{| #1 \rangle}

\newcommand{\braket}[1]{\langle #1 \rangle}

\begin{document}
\title{Theory of high-gain twin-beam generation in waveguides: \\ from Maxwell's equations to efficient simulation}
\author{Nicol\'as Quesada\footnote{Equal contributors}\footnote{Current affiliation: Xanadu, Toronto, Canada } }
\affiliation{Perimeter Institute for Theoretical Physics, Waterloo, ON, N2L 2Y5,
Canada}
\author{Gil Triginer$^{*}$}
\affiliation{Clarendon Labs, Department of Physics, Oxford University, Parks Road
OX1 3PU Oxford}
\author{Mihai D. Vidrighin$^{*}$}
\affiliation{Clarendon Labs, Department of Physics, Oxford University, Parks Road
OX1 3PU Oxford}
\author{J.E. Sipe}
\affiliation{Department of Physics, University of Toronto, Toronto, ON, M5S 1A7,
Canada}
\begin{abstract}
We provide an efficient method for the calculation of high-gain, twin-beam generation in waveguides derived from a canonical treatment of Maxwell's equations.  Equations of motion are derived that naturally accommodate photon generation via spontaneous parametric down-conversion (SPDC) or spontaneous four-wave mixing (SFWM), and also include the effects of both self-phase modulation (SPM) of the pump, and of cross-phase modulation(XPM) of the twin beams by the pump.  The equations we solve involve fields that evolve in space and are labelled by a frequency.  We provide a proof that these fields satisfy bonafide commutation relations, and that in the distant past and future they reduce to standard time-evolving Heisenberg operators. Having solved for the input-output relations of these Heisenberg operators we also show how to construct the ket describing the quantum state of the twin-beams. Finally, we consider the example of high-gain SPDC in a waveguide with a flat nonlinearity profile, for which our approach provides an explicit solution that requires only a single matrix exponentiation.
\end{abstract}
\maketitle

\section{Introduction}

The generation of twin beams is an important technique for the
production of nonclassical light \cite{andersen201630}. In  early
experiments , the twin beams were generated over a manifold of modes.
This was because the nonlinear medium was pumped with a quasi-continuous-wave
source. As pulsed sources were developed and mode engineering
improved, it became possible to drastically reduce the number of spatio-temporal
modes to essentially just one \cite{mosley2008heralded}. Furthermore,
recent developments in photonics have allowed for the tight confinement
of the travelling waves participating in the three- or four-wave mixing
process necessary for the generation of twin beams \cite{harder2013optimized,finger2015raman,harder2016single}.
These developments have moved the focus of theoretical descriptions
of twin beam generation from the perturbative regime to the nonperturbative
regime.

Theoretical descriptions of twin beam generation broadly follow three
approaches, each one of which can be identified by the spacetime variables
  used to describe the propagation of states, Heisenberg operators,
or their correlation functions. The first is a $(\vec{k},t)$ approach
\cite{liscidini2012asymptotic,yang2008spontaneous,quesada2017effects},
in which the amplitudes of expansion fields specified by wave vectors
$\vec{k}$ are propagated in time. As the vectorial nature of $\vec{k}$
suggests, this strategy can be applied to propagation geometries in
any number of dimensions. It has not yet been extended beyond the
perturbative regime.

The second is a $(z,t)$ approach, in which slowly varying envelope
operators are propagated forward in time \cite{lai1995general,reddy2017engineering,reddy2017temporal}.
This strategy can accommodate dispersion, but it requires the calculation
of the propagation of a sufficiently complete set of classical pulses
undergoing the nonlinear dynamics of a stimulated experiment, and
then the use of this information to describe the spontaneous experiment. For certain limiting situations, no numerics are needed since the equations of motion admit an analytic solution \cite{mckinstrie2017single}.

The third strategy is a $(z,\omega)$ approach, where one deals with
Fourier transforms of the $(z,t)$ operators \cite{klyshko2018photons,kolobov1999spatial, christ2013theory,lipfert2018bloch,mauerer2009colours}.
This approach has been heavily used since the early days of quantum
nonlinear optics, and has been justified, \emph{e.g., }\textit{\emph{by}}
Bergman\cite{bergman1994quantum}, who argued that ``Evolution in
time of an operator in the Heisenberg picture is given by its commutation
with the Hamiltonian. Here the propagation distance, $z$, plays the
role of time.'' However, Huttner \emph{et al.} \ \cite{huttner1992quantization}
pointed out that this approach ``is not derived from a Lagrangian
and therefore has not been justified in terms of a canonical scheme.''
As noted by Haus \cite{shirasaki1990squeezing,haus1984waves}, the
validity of the argument expressed by Bergman and used by many others
arises physically because ``the formalism implies the application
to narrowband spectra within which such a frequency independence can
be assumed and a group velocity defined.'' In even simpler terms:
\emph{if a group velocity $v$ can be defined, then time =position/$v$}. 

In this paper, we focus on the regime where such a simple link between space and time is provided by a group velocity.
We provide a rigorous proof of the validity of the
$(z,\omega)$ approach for twin beam generation, connect it to canonical
(Hamiltonian) schemes, and use it to study twin-beam generation via
SPDC or SFWM in the high gain regime. We do this by showing that,
even in the presence of a nonlinear medium, suitably defined field
operators $a(z,\omega)$ satisfy correct commutation relations \emph{if}
the dispersion relation for the mode specified by the operator $a(z,\omega)$
is linear,  $k(\omega)=k(\bar{\omega})+(\omega-\bar{\omega})/v$
where $\bar{\omega}$ is some properly defined central frequency.
Furthermore, we show that, if the relation between the wavevector
and the frequency is \textit{not} linear (in the simplest case quadratic,
as for example as considered by Caves and Crouch  \cite{caves1987quantum}),
then the field operators $a(z,\omega)$ defined here for the twin beams have pathological commutation relations. 

To derive these results, in Sec. \ref{sec:quant} we provide a self-contained
derivation of the equations of motion of the quantum operators that
classically correspond to slowly varying envelope functions, starting
from Maxwell's equations and a Hamiltonian canonical quantization
procedure \cite{huttner1992quantization,born34,hillery85,abram1987quantum,bhat2006hamiltonian,drummond14,quesada2017you,kennedy1988quantization}.
In Sec. \ref{sec:dynamics}, we introduce the $(z,\omega)$ operators,
which are the Fourier transforms of the $(z,t)$ operators, and derive
their equations of motion. These equations account for twin beam generation
via SPDC or SFWM, and also include automatically phase-matched interactions
such as self-phase modulation (SPM) of the pump, and cross-phase modulation
(XPM) of the generated twin beams by the pump. In Sec. \ref{soleoms}
we show that these equations, upon discretization, can be efficiently
solved using matrix exponentiation, and study some properties of their
solution by the introduction of Schmidt modes. In Sec. \ref{sec:example},
we use the techniques developed in the previous sections to study
spontaneous twin beam generation, and as an example consider
a homogeneous medium with a pump beam that does not undergo SPM. Under
these circumstances, the solution of the equations of motion can be
reduced to a single matrix exponentiation. In a companion
paper \cite{triginer2018complete} we use these techniques to validate
a recent tomographical method for the characterization of two-mode
squeezing in the high-gain regime. Finally, in Sec. \ref{sec:conc} we
present some general conclusions and comment on the validity of the
$(z,\omega)$ approach when the relation between $k$ and $\omega$
cannot be approximated by a linear function; a detailed calculation
is presented in Appendix \ref{app:disp}.

\section{Quantization in nonlinear media}

\label{sec:quant} In this section we quantize the electromagnetic
field in a source-free nonlinear material, and obtain
the Hamiltonian governing the generation of photons in twin beams
via SPDC/SFWM, the self-phase modulation of the pump, and the cross
phase modulation of the twin beams by the pump.

\subsection{Quantization}

We start by writing Maxwell's equations in a source-free
medium \begin{subequations} \label{MEs} 
\begin{align}
\frac{\partial}{\partial t}\mathbf{B} & =-\mathbf{\nabla}\times\mathbf{E},\label{faraday}\\
\frac{\partial}{\partial t}\mathbf{D} & =\mathbf{\nabla}\times(\mathbf{B}/\mu_{0}),\label{ampere}\\
\mathbf{\nabla}\cdot\mathbf{D} & =\mathbf{\nabla}\cdot\mathbf{B}=0.\label{gauss}
\end{align}
\end{subequations} We take $\mathbf{B}$ and $\mathbf{D}$ as the
fundamental fields \cite{born34,hillery85,abram1987quantum,bhat2006hamiltonian,drummond14,quesada2017you},
and write the polarization appearing in the constitutive relation,
\begin{align}
\mathbf{P}=\mathbf{D}-\epsilon_{0}\mathbf{E},\label{const}
\end{align}
as a function of the \emph{displacement field} $\mathbf{D}$, 
\begin{align}
\mathbf{P}(\mathbf{D})=\Gamma^{(1)}\mathbf{D}+\Gamma^{(2)}\mathbf{D}^{2}+\Gamma^{(3)}\mathbf{D}^{3}+\ldots
\end{align}
 The notation here is schematic, but of course indicates the appropriate
summation over Cartesian components; for the moment we neglect any
material dispersion. Having expressed the macroscopic polarization
in terms of $\mathbf{D}$, we can now write the energy density of
the system as 
\begin{align}
\mathcal{H} & =\int\mathbf{E}(\mathbf{D})\cdot d\mathbf{D}+\int\mathbf{H}(\mathbf{B})\cdot d\mathbf{B}\\
 & =\frac{\mathbf{B}^{2}}{2\mu_{0}}+\frac{1-\Gamma^{(1)}}{2\epsilon_{0}}\mathbf{D}^{2}-\frac{\Gamma^{(2)}\mathbf{D}^{3}}{3\epsilon_{0}}-\frac{\Gamma^{(3)}\mathbf{D}^{4}}{4\epsilon_{0}}-\ldots,\label{eq:Hamiltonian_density}
\end{align}
with the Hamiltonian $H$ given by the integral over space of this
density. The Heisenberg equations of motion, which for an arbitrary
operator are 
\begin{align}
i\hbar\frac{d}{dt}O(t)=[O(t),H],\label{heo}
\end{align}
give precisely Maxwell's equations (\ref{faraday},\ref{ampere})
for the operators $\mathbf{D}$ and $\mathbf{B}$ if one uses the
Hamiltonian $H$ defined above and the commutation relations \cite{born34,volkov2004nonlinear}
\begin{subequations}\label{CR} 
\begin{align}
[D_{j}(\mathbf{r}),B_{l}(\mathbf{r}')] & =i\hbar\varepsilon_{jlm}\frac{\partial}{\partial r_{m}}\delta(\mathbf{r}-\mathbf{r}'),\\{}
[D_{j}(\mathbf{r}),D_{l}(\mathbf{r}')] & =[B_{j}(\mathbf{r}),B_{l}(\mathbf{r}')]=0.
\end{align}
\end{subequations} In Eq. (\ref{CR}) the indices $j,l,m$ denote
Cartesian components, $\varepsilon_{jlm}$ is the Levi-Civita symbol
and $\delta(\mathbf{r})$ is the Dirac distribution. The divergence
conditions (\ref{gauss}) are satisfied by choosing a basis of modes
that are divergenceless; see Eq. (\ref{eq:Bdivless},\ref{eq:Ddivless})
below. Note that if instead one quantized in terms $\mathbf{E}$ and
$\mathbf{B}$ one would not obtain Maxwell's equations (\ref{faraday},\ref{ampere})
as the Heisenberg equations of motion for such fields \cite{quesada2017you}.
Furthermore, note that $\mathbf{D}$ and $\mathbf{B}$ are transverse,
unlike $\mathbf{E}$.

\begin{figure}
\includegraphics[width=0.95\linewidth]{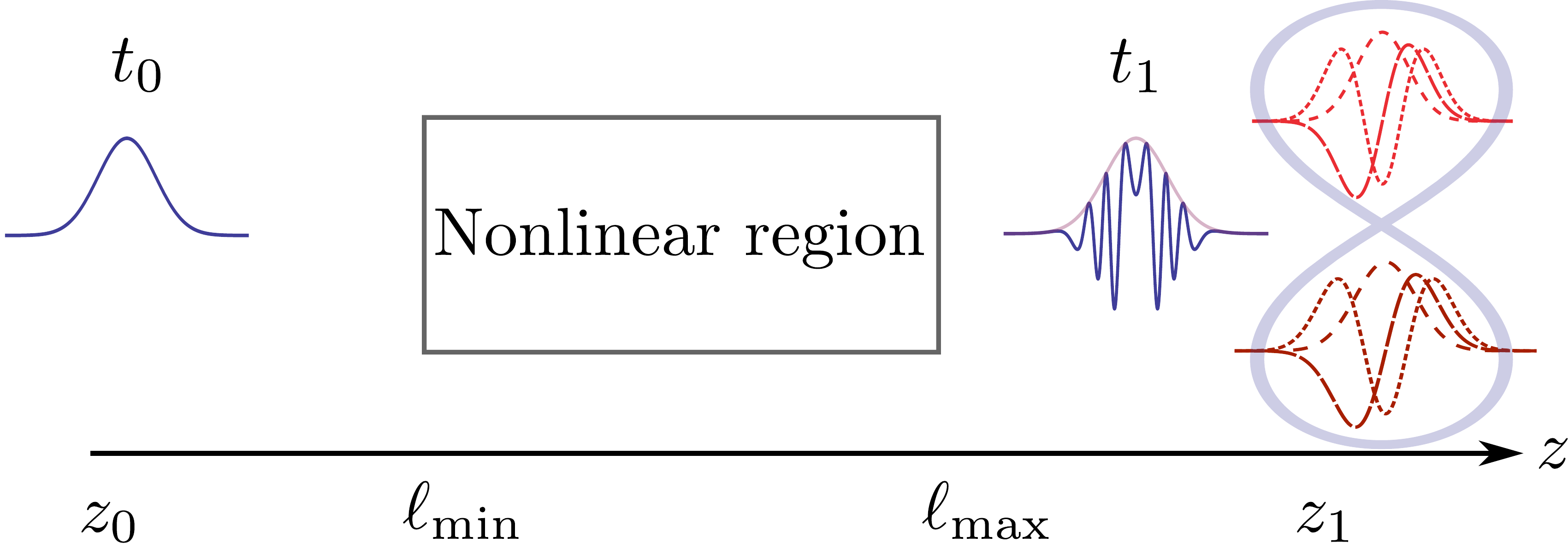}
\caption{\label{geometry1D} Propagation geometry. A pump field localized around
$z_{0}$ is directed towards the nonlinear region, where $z\in[\ell_{\min},\ell_{\max}]$.
After the pump field has left the nonlinear region it has undergone
self-phase modulation and has created twin-beams in a set of Schmidt
modes, indicated by the dashed waveforms in the right hand side of
the figure.}
\end{figure}

\subsection{Linear Fields Expansion}

To introduce expansion fields for the displacement and magnetic fields
we follow the approach of Bhat and Sipe \cite{bhat2006hamiltonian}.
This approach can be generalized to include material dispersion in
the linear response of the medium; we simply sketch the results. We
consider fields in the linear regime of the form $f(\mathbf{r},t)=f_{\mu k}(\mathbf{r})\exp(-i\omega_{\mu k}t)+$c.c.
They will satisfy the linear Maxwell equations if they satisfy the
so-called master equation\cite{joan11} 
\begin{align}
 & \mathbf{\nabla\times}\left[\frac{\mathbf{\nabla\times B}_{\mu k}(\mathbf{r})}{n^{2}(x,y;\omega_{\mu k})}\right]=\left(\frac{\omega_{\mu k}}{c}\right)^{2}\mathbf{B}_{\mu k}(\mathbf{r}),\label{master}
\end{align}
and also 
\begin{align}
\mathbf{\nabla}\cdot\mathbf{B}_{\mu k}(\mathbf{r}) & =0,\label{eq:Bdivless}\\
\mathbf{D}_{\mu k}(\mathbf{r}) & =\frac{i}{\mu_{0}\omega_{\mu k}}\mathbf{\nabla\times B}_{\mu k}(\mathbf{r}),\label{eq:Ddivless}
\end{align}
where $n(x,y;\omega)$ is the (local) position and frequency dependent
refractive index. In the nondispersive limit and for an isotropic
material, the $\Gamma^{(1)}$ coefficient can be related to the more
standard linear polarizability $\chi^{(1)}$ and the index of refraction
$n$ as follows: 
\begin{align}
1-\Gamma^{(1)}=\left(1+\chi^{(1)}\right)^{-1}=\frac{1}{n^{2}}.
\end{align}

We take the refractive index to be independent of $z$, the distance
along a waveguide. Then the solution of the master equation will be
of the form 
\begin{align}
\mathbf{D}_{\mu k}(\mathbf{r})=\frac{\mathbf{d}_{\mu k}(x,y)}{\sqrt{2\pi}}e^{ikz},\quad\mathbf{B}_{\mu k}(\mathbf{r})=\frac{\mathbf{b}_{\mu k}(x,y)}{\sqrt{2\pi}}e^{ikz},\label{eq:expansion}
\end{align}
where the label $k$ is a wavevector, and we use a Greek label
$\mu$ to identify which field we are describing, writing $\mu=p$
for the pump and $\mu=s,i$ for the twin beam fields.

This is a convenient expansion basis for the field operators $\mathbf{D}(\mathbf{r},t)$
and $\mathbf{B}(\mathbf{r},t)$ even in the presence of material dispersion,
under the assumption that at frequencies of interest there is no absorption;
 normalization must then be  done according to  
\begin{align}
\int dxdy\frac{\mathbf{d}_{\mu k}^{\ast}(x,y)\cdot\mathbf{d}_{\mu k}(x,y)}{\epsilon_{0}n^{2}(x,y;\omega_{\mu k})}\frac{v_{\text{ph}}(x,y;\omega_{\mu k})}{v_{\text{g}}(x,y;\omega_{\mu k})}=1,\label{norm}
\end{align}
where $v_{\text{ph}}(x,y;\omega)$ and $v_{\text{g}}(x,y;\omega)$
are respectively the local phase and group velocities at each point
in the waveguide \cite{bhat2006hamiltonian}.

A rough estimate of the magnitude of these  coefficients can be obtained
by assuming that the field has a transverse area $A$, giving 
\begin{align}
|\mathbf{d}|\approx\sqrt{\frac{\epsilon_{0}n^{2}v_{g}}{v_{p}A}},
\end{align}
and we assume that the index of refraction and group and phase velocities
are evaluated at some central frequency of interest. Using the fields
in Eq. (\ref{eq:expansion}) as basis functions normalized according
to Eq. (\ref{norm}), the displacement and magnetic fields can be
written in the following very symmetric form 
\begin{align}
 & \mathbf{B}(\mathbf{r})=\sum_{\mu}\int dk\sqrt{\frac{\hbar\omega_{\mu k}}{2}}b_{\mu k}\mathbf{B}_{\mu k}(\mathbf{r})+\text{H.c.},\label{eq:Bexpand}\\
 & \mathbf{D}(\mathbf{r})=\sum_{\mu}\int dk\sqrt{\frac{\hbar\omega_{\mu k}}{2}}b_{\mu k}\mathbf{D}_{\mu k}(\mathbf{r})+\text{H.c.},\label{eq:Dexpand}
\end{align}
and furthermore, the linear part of the Hamiltonian can then be written
as 
\begin{align}
 & H_{L}=\int dk\sum_{\mu}\hbar\omega_{\mu k}b_{\mu k}^{\dagger}b_{\mu k},\label{eq:DandB-1}
\end{align}
with the neglect of zero-point energy.

The creation and destruction operators $b_{\mu k}^{\dagger}$ and
$b_{\mu k}$ satisfy the bosonic commutation relations \cite{bhat2006hamiltonian}
\begin{align}
[b_{\mu k},b_{\mu'k'}] & =[b_{\mu k}^{\dagger},b_{\mu'k'}^{\dagger}]=0,\\{}
[b_{\mu k},b_{\mu'k'}^{\dagger}] & =\delta_{\mu\mu'}\delta(k-k'),
\end{align}
where recall we use the Greek label $\mu\in\{p,s,i\}$ to refer to
the three fields of interest pump, signal and idler.

At this point the index $\mu$
is superfluous if the pump, signal, and idler expansion fields are associated
with the same transverse profile function in the $xy$ plane. This is often
true for SFWM, but not for SPDC. We henceforth redefine the index $\mu$ to
indicate both the different ranges of $k$ associated with the pump, signal,
and idler, and their transverse profile functions. We now introduce
\emph{field} operators 
\begin{align}
\psi_{\mu}(z)=\int\frac{dk}{\sqrt{2\pi}}b_{\mu k}e^{i(k-\bar{k}_{\mu})z}.\label{psidef}
\end{align}
which are quantum operators analogous to the slowly varying envelope
functions in space, since we have removed a central wavevector $\bar{k}_{\mu}$
associated with the central frequency $\bar{\omega}_{\mu}$. In the
limit where group velocity dispersion in each field can be neglected,
the dispersion relation for each field, with group velocity $v_{\mu}$,
can be written as 
\begin{align}
k-\bar{k}_{\mu}=(\omega-\bar{\omega}_{\mu})/v_{\mu}.\label{linearDisp}
\end{align}
The Schr\"odinger picture field operators satisfy the commutation relations
\begin{align}
[\psi_{\mu}(z),\psi_{\mu'}(z')] & =[\psi_{\mu}^{\dagger}(z),\psi_{\mu'}^{\dagger}(z')]=0,\label{scheqt}\\{}
[\psi_{\mu}(z),\psi_{\mu'}^{\dagger}(z')] & =\delta_{\mu,\mu'}\delta(z-z'),
\end{align}
again, under the assumptions that the pump, signal, and idler fields
span different wavevector and frequency ranges, and thus that for
each field operator (\ref{psidef}) we can formally let $k$ range
from $-\infty$ to $\infty$ when evaluating the commutation relations.

Now we assume that group velocity does not vary significantly over
the bandwidths of interest, ignoring group velocity dispersion. Then
the linear part of the Hamiltonian, given in Eq. (\ref{eq:DandB-1}),
can be written   as
\begin{align}
H_{\text{L}}= & \sum_{\mu}\hbar\bar{\omega}_{\mu}\int dz\;\psi_{\mu}^{\dagger}(z)\psi_{\mu}(z)\label{Schroedinger1}\\
 & +\frac{i}{2}\sum_{\mu}\hbar v_{\mu}\int dz\left(\frac{\partial\psi_{\mu}^{\dagger}(z)}{\partial z}\psi_{\mu}(z)-\psi_{\mu}^{\dagger}(z)\frac{\partial\psi_{\mu}(z)}{\partial z}\right),\nonumber 
\end{align}
(see Appendix \ref{app:linearpsi}). The second term on the last equation
accounts for the linear dependence of the frequency on momentum in
reciprocal space, which in real space acts as a derivative on the
field operator.

We can write the displacement field $\mathbf{D(}\mathbf{r})$ (\ref{eq:Dexpand})
in terms of the field operators as 
\begin{align}
 & \mathbf{D}(\mathbf{r})\approx\sum_{\mu}e^{i\bar{k}_{\mu}z}\left[\sqrt{\frac{\hbar\omega_{\mu}}{2}}\mathbf{d}_{\mu{k}_{\mu}}(x,y)\right]_{k_{\mu}=\bar{k}_{\mu}}\psi_{\mu}(z)+\text{H.c.},\label{eq:Dexpan}
\end{align}
where we have performed a Taylor expansion of the terms inside the integral
around $k_{\mu}=\bar{k}_{\mu}$, and assumed any variation in the transverse
mode profiles $\mathbf{d}_{\mu k_{\mu}}$ and the frequencies $\omega_{\mu}$
to be negligible; the magnetic field $\mathbf{B}(\mathbf{r})$ (\ref{eq:Bexpand})
can be written in a similar way.

\subsection{The Nonlinear Interaction}

We now turn to the nonlinear part of the Hamiltonian, which is given
by the integral over space of the third and fourth terms on the right-hand-side
of (\ref{eq:Hamiltonian_density}). Explicitly indicating Cartesian
components and with the usual Einstein summation convention we have
\begin{align}
H_{\text{NL}}= & -\frac{1}{3\epsilon_{0}}\int d\mathbf{r}\ \Gamma_{2}^{ijl}(\mathbf{r})\ D^{i}(\mathbf{r})D^{j}(\mathbf{r})D^{l}(\mathbf{r})\label{HNL}\\
 & -\frac{1}{4\epsilon_{0}}\int d\mathbf{r}\ \Gamma_{3}^{ijlm}(\mathbf{r})\ D^{i}(\mathbf{r})D^{j}(\mathbf{r})D^{l}(\mathbf{r})D^{m}(\mathbf{r}).\nonumber 
\end{align}
In terms of the usual nonlinear susceptibilities $\chi_{2}^{ijl}(x,y,z)$
and $\chi_{3}^{ijlm}(x,y,z)$ characterizing the second and third
order optical response, we have
\begin{align}
\Gamma_{2}^{ijl}(x,y,z)= & \frac{\chi_{2}^{ijl}(x,y,z)}{\epsilon_{0}n_{o}^{6}(x,y)},\\
\Gamma_{3}^{ijlm}(x,y,z)= & \frac{\chi_{3}^{ijlm}(x,y,z)}{\epsilon_{0}^{2}n_{o}^{8}(x,y)}\\
 & -\sum_{q}\frac{2\chi_{2}^{ijq}(x,y,z)\chi_{2}^{qlm}(x,y,z)/n_{q}^{2}(x,y)}{\epsilon_{0}^{2}n_{o}^{8}(x,y)},\nonumber 
\end{align}
where we neglect the effects of material dispersion on the nonlinear
Hamiltonian, and take $n_{o}(x,y)$ to be an index of refraction at
some ``typical'' wavelength \cite{volkov2004nonlinear} . We can
now write the nonlinear Hamiltonian Eq. (\ref{HNL}) in terms of the
field operators $\psi_{\mu}(z)$, considering processes in which three
beams labelled pump ($p$), signal ($s$), and idler ($i$) are coupled
by the nonlinear interaction. We assume we can choose our centre frequencies
$\bar{\omega}_{\mu}$ and the associated wave vectors $\bar{k}_{\mu}$
such that either \begin{subequations} \label{PM} 
\begin{align}
2\bar{\omega}_{p}-\bar{\omega}_{s}-\bar{\omega}_{i} & =0,\\
2\bar{k}_{p}-\bar{k}_{s}-\bar{k}_{i} & =0,
\end{align}
\end{subequations} or \begin{subequations} \label{PM3} 
\begin{align}
\bar{\omega}_{p}-\bar{\omega}_{s}-\bar{\omega}_{i} & =0,\\
\bar{k}_{p}-\bar{k}_{s}-\bar{k}_{i} & =0.\label{pm3}
\end{align}
\end{subequations} The first condition will allow for the creation
of twin beams via spontaneous four wave mixing (SFWM) and the second
condition will allow for their creation via spontaneous parametric
down-conversion (SPDC). Note that both conditions cannot be satisfied
at the same time. Yet even if only the SPDC process is phase matched,
other third-order nonlinear processes, such as self- and cross-phase
modulation, are still phase matched, and can modify the properties
of the photons generated in SPDC. Of course, this will also happen
if SFWM is used to generate photons instead of SPDC . Note that if
\textit{quasi}-phase matching is used for a second order process,
the RHS of Eq. (\ref{pm3}) should be changed to $\pm2\pi/\Lambda_{\text{pol}}$
where $\Lambda_{\text{pol}}$ is the poling period.

Under these assumptions we can write the nonlinear part of the Hamiltonian
as \begin{subequations}\label{nonlin} 
\begin{align}
H_{\text{NL}}= & -\hbar\int dz\Bigg\{\frac{1}{2}\ \zeta_{p}(z)\psi_{p}^{\dag}(z)\psi_{p}^{\dag}(z)\psi_{p}(z)\psi_{p}(z)\Bigg.\label{SPM1}\\
 & +\zeta_{i}(z)\psi_{p}^{\dagger}(z)\psi_{p}(z)\psi_{i}^{\dag}(z)\psi_{i}(z)\label{XPM1}\\
 & +\zeta_{s}(z)\psi_{p}^{\dagger}(z)\psi_{p}(z)\psi_{s}^{\dag}(z)\psi_{s}(z)\label{XPM2}\\
 & +\left(\xi_{\delta}(z)\psi_{s}^{\dagger}(z)\psi_{i}^{\dagger}(z)\left(\psi_{p}(z)\right)^{\delta}+\text{H.c.}\right)\Bigg.\Bigg\},\label{SFWM}
\end{align}
\end{subequations} where we assume full permutation symmetry of
the $\Gamma$ tensors in their Cartesian indices, and keep  the terms
that are energy and phase matched consistent with Eqs. (\ref{PM})
and (\ref{PM3}); we have also introduced the quantities $\zeta_{p},\zeta_{i},\zeta_{s},\xi_{\delta}$,
defined in detail in Appendix \ref{coeff:app}, which capture the
strength of the nonlinear interactions corresponding to SPM of the
pump (\ref{SPM1}), XPM between the pump and the idler (\ref{XPM1}),
XPM between the pump the signal (\ref{XPM2}) and twin-beam generation
via either SPDC ($\delta=1$) or SFWM ($\delta=2$) (\ref{SFWM})
respectively. We take these quantities to be nonzero only in a region
$\ell_{\min}\leq z\leq\ell_{\max}$ where the nonlinear coupling occurs;
this is schematically represented in Fig. \ref{geometry1D}. Notice
that in the last set of equations we have only included SPM of the
pump, since the intensities of the signal and idler field are typically small
enough for SPM to be negligible.

\section{Dynamics of the fields}

\label{sec:dynamics} With the full Hamiltonian of the system in place
we can write the Schr\"{o}dinger equation satisfied by the evolution operator,
\begin{align}
i\hbar\frac{d}{dt}\mathcal{\hat{U}}(t,t_{0})=\left(H_{\text{L}}+H_{\text{NL}}\right)\mathcal{\hat{U}}(t,t_{0}),
\end{align}
where $t_{0}$ is conventionally the time at which the Schr\"odinger
and Heisenberg pictures coincide; $\mathcal{\hat{U}}(t_{0},t_{0})=\mathbb{I}$,
where $\mathbb{I}$ is the identity operator. We take this time to
be long before the pump beam enters the nonlinear region. Once the
unitary evolution operator is obtained one can propagate the operators,
for example, 
\begin{align}
b_{\mu k}(t_{1}) & =\mathcal{\hat{U}}^{\dagger}(t_{1},t_{0})b_{\mu k}(t_{0})\mathcal{\hat{U}}(t_{1},t_{0})\label{haction}\\
 & =\mathcal{F}(b_{\mu'k'}(t_{0}),b_{\mu'k'}^{\dagger}(t_{0})),\nonumber 
\end{align}
In the last equation we use $\mathcal{F}$ to indicate that the quantities
on the left hand side, the operators at time $t_{1}$, are functions
of all the operators at $t_{0}$.

The main objective of the next sections will be to provide a detailed
derivation of the mapping connecting time evolving operators at some
$t=t_{0}$ in the distant past with operators at $t=t_{1}$ in the
\emph{distant future}, long after the pump pulse has exited the nonlinear
region. Henceforth we assume $t_{0}$ and $t_{1}$ are so chosen.

\subsection{Pump dynamics}

We first look at the Heisenberg equation of motion for the pump field,
which follows from using the commutation relations (\ref{scheqt})
with a Hamiltonian that is the sum of the linear (\ref{Schroedinger1})
and nonlinear (\ref{nonlin}) contributions, 
\begin{align}
 & \left(\frac{\partial}{\partial t}+v_{p}\frac{\partial}{\partial z}+i\bar{\omega}_{p}\right)\psi_{p}(z,t)=\label{SPM}\\
 & \quad i\zeta_{p}(z)\psi_{p}^{\dagger}(z,t)\psi_{p}(z,t)\psi_{p}(z,t)+\text{back-action terms},\nonumber 
\end{align}
where the ``back-action terms'' are contributions that contain the
operators $\psi_{s}(z,t)$ and $\psi_{i}(z,t)$. We assume that the
pump field is prepared in a strong coherent state with a large number
of photons, and we assume that this number remains unchanged during
the SFWM or SPDC process; we may then ignore the back-action terms,
which are all proportional to the first power of $\psi_{p}(z,t)$
and second powers of $\psi_{s}(z,t)$ and $\psi_{i}(z,t)$, and have
a much smaller effect than the self-phase modulation term appearing
in the right hand side of Eq. (\ref{SPM}). Furthermore, because of
the undepleted-classical pump approximation just described we replace $\psi_{p}(z,t)\to\braket{\psi_{p}(z,t)}$.
The solution to the equation of motion for the pump mean field is
\begin{align}
\braket{\psi_{p}(z,t)}= & \Lambda(z-v_{p}(t-t_{0})) e^{-i \bar{\omega}_{p}(t-t_{0}) + i  \varphi(z,t)},
\label{pumpSol}
\end{align}
where the phase accumulated due to SPM is 
\begin{align}
\varphi(z,t)= & |\Lambda(z-v_{p}(t-t_{0}))|^{2}\int_{t_{0}}^{t}dt'\zeta_{p}(z-v_{p}(t-t')),\ 
\end{align}
and where we introduced 
\begin{align}
\braket{\psi_{p}(z,t_{0})} & =\Lambda(z).
\end{align}
The mean number of photons in the pump pulse is 
\begin{align}
N_{p}=\int dz|\braket{\psi_{p}(z,t)}|^{2}=\int dz|\Lambda(z)|^{2}\gg 1
\end{align}
and its energy is simply $\mathcal{E}_{p}=\hbar\bar{\omega}_{p}N_{p}$.
The spatial distribution of the energy in the field will not be affected by SPM, 
\begin{align}
|\braket{\psi_{p}(z,t)}|^{2}=|\Lambda(z-v_{p}(t-t_{0}))|^{2},\label{simplemod2}
\end{align}
and thus the spectral content (i.e. the Fourier transform) of $|\braket{\psi_{p}(z,t)}|^{2}$
remains unchanged under propagation; see Appendix \ref{ders:app}
for details.

\subsection{Twin-beam dynamics}

We can now calculate the Heisenberg equations of motion for the signal
and idler field operators $\psi_{s},\psi_{i}^{\dagger}$, \begin{subequations}
\label{eomszt} 
\begin{align}
\left(\frac{\partial}{\partial t}+v_{s}\frac{\partial}{\partial z}+i\bar{\omega}_{s}\right) & \psi_{s}(z,t)=\label{psi1Heisenberg}\\
 & i\xi_{\delta}(z)\braket{\psi_{p}(z,t)}^{\delta}\psi_{i}^{\dagger}(z,t)\nonumber \\
 & +i\zeta_{s}(z)|\braket{\psi_{p}(z,t)}|^{2}\psi_{s}(z,t),\nonumber \\
\left(\frac{\partial}{\partial t}+v_{i}\frac{\partial}{\partial z}-i\bar{\omega}_{i}\right) & \psi_{i}^{\dagger}(z,t)=\label{psi2Heisenberg}\\
 & -i\xi_{\delta}^{*}(z)\braket{\psi_{p}^{\dagger}(z,t)}^{\delta}\psi_{s}(z,t)\nonumber \\
 & -i\zeta_{i}(z)|\braket{\psi_{p}(z,t)}|^{2}\psi_{i}^{\dagger}(z,t).\nonumber 
\end{align}
\end{subequations} The right hand sides of Eqs. (\ref{eomszt}a,\ref{eomszt}b)
for $\psi_{s}$ and $\psi_{i}^{\dagger}$ account for photon generation
via either SPDC ($\delta=1$) or SFWM ($\delta=2$), and for cross-phase
modulation of the pump on the signal and idler fields . The left hand
side in Eqs. (\ref{eomszt}) accounts for propagation at group velocity
$v_{j}$, and oscillation at frequency $\bar{\omega}_{j}$. If group
velocity dispersion were included within the bandwidth of each field,
then further terms proportional to $\partial^{2}\psi_{s,i}/\partial z^{2}$
would also be present.

Henceforth we neglect group velocity dispersion within each of the
pump, signal, and idler bandwidths, and introduce the following operators
for the signal and idler fields 
\begin{align}
a_{j}(z,\omega) & =\int\frac{dt}{\sqrt{2\pi/v_{j}}}e^{i(\omega t-z(\omega-\bar{\omega}_{j})/v_{p})}\psi_{j}(z,t),\\
\psi_{j}(z,t) & =\int\frac{d\omega}{\sqrt{2\pi v_{j}}}e^{-i(\omega t-z(\omega-\bar{\omega}_{j})/v_{p})}a_{j}(z,\omega),
\end{align}
where in the last set of equations we used the Latin label $j\in\{s,i\}$
exclusively to refer to the twin beams, signal and idler, and omitting
the pump. The fields $a_{j}(z,\omega)$ are the $(t,\omega)$ Fourier
transforms of the slowly varying envelope field operators in a moving
frame at the group velocity of the pump field $v_{p}$ \cite{vidrighin2017quantum}.
The equations for the spatial evolution of the $a_{j}(z,\omega)$
are then found to be (see Appendix \ref{ders:app} for a derivation)
\begin{subequations}\label{eomszw} 
\begin{align}
\frac{\partial}{\partial z} & a_{s}(z,\omega)=i\Delta k_{s}(\omega)a_{s}(z,\omega)\\
 & +i\frac{\gamma_{\text{XPM},s}h_{s}(z)}{2\pi}\int d\omega' \mathcal{E}_p(\omega-\omega')a_{s}(z,\omega')\nonumber \\
 & +i\frac{\gamma_{\delta}g(z)}{\sqrt{2\pi}}\int d\omega'\beta_{p}(z,\omega+\omega')a_{i}^{\dagger}(z,\omega'),\nonumber \\
\frac{\partial}{\partial z} & a_{i}^{\dagger}(z,\omega)=-i\Delta k_{i}(\omega)a_{i}(z,\omega)\\
 & -i\frac{\gamma_{\text{XPM},i}h_{i}(z)}{2\pi}\int d\omega'\mathcal{E}_p^{*}(\omega-\omega')a_{i}^{\dagger}(z,\omega')\nonumber \\
 & -i\frac{\gamma_{\delta}^{*}g(z)}{\sqrt{2\pi}}\int d\omega'\beta_{p}^{*}(z,\omega+\omega')a_{s}(z,\omega'),\nonumber 
\end{align}
\end{subequations} The first term on the right hand side of these
equations  describes the pulse walk-off between the pump and the signal/idler;
 we have defined 
\begin{equation}
\Delta k_{j}(\omega)=\left(\frac{1}{v_{j}}-\frac{1}{v_{p}}\right)\left(\omega-\bar{\omega}_{j}\right)\ .
\end{equation}
The second term, accounting for cross-phase modulation, contains a
coupling strength profile, 
\begin{equation}
\gamma_{\text{XPM},j}h_{j}(z)=\frac{\zeta_{j}(z)}{v_{p}v_{j}\hbar\bar{\omega}_{p}},
\end{equation}
where we take $h_{j}(z)=1,0$ respectively in the region where the
nonlinearity is present or absent, and the $(t,\omega)$ Fourier transform
of the energy distribution of the field in the moving frame is 
\begin{equation}
\mathcal{E}_p(\omega)=\mathcal{E}_p^{*}(-\omega)=e^{i\omega t_{0}}\hbar\bar{\omega}_{p}\int dz|\Lambda(z)|^{2}e^{-i\omega z/v_{p}}.
\end{equation}
The last term is responsible for twin-beam generation, and contains
a coupling strength profile 
\begin{equation}
\gamma_{\delta}g(z)=\frac{\xi_{\delta}(z)}{\sqrt{v_{p}v_{s}v_{i}(\hbar\bar{\omega}_{p})^{\delta}}}
\end{equation}
with $g(z)=0$ where the nonlinearity is absent and either $1$ or
$-1$ (the latter to describe quasi-phase matching) where the nonlinearity
is present, and the $(t,\omega)$ Fourier transform of the pump amplitude
in the moving frame is

\begin{align}
 & \beta_{p}(z,\omega)=\frac{(\hbar\bar{\omega}_{p})^{\delta/2}}{\sqrt{2\pi/v_{p}}}\int dt\ e^{i(\omega t-z(\omega-\delta\bar{\omega}_{p})/v_{p})}\braket{\psi_{p}(z,t)}^{\delta}\nonumber \\
 & =e^{i\omega t_{0}}\frac{(\hbar\bar{\omega}_{p})^{\delta/2}}{\sqrt{2\pi v_{p}}}\int dz'e^{-iz'\frac{(\omega-\bar{\omega}_{p}\delta)}{v_{p}}}\left(\Lambda(z')\right)^{\delta}e^{i\delta\ \theta(z,z'))}\label{pumpspect}
\end{align}

with a nonlinear phase

\begin{align}
 & \theta(z,z')\equiv\varphi\left(z,t_{0}+\tfrac{z-z'}{v_{p}}\right)=|\Lambda(z')|^{2}\int_{z'}^{z}\frac{dz''}{v_{p}}\zeta_{p}(z'').
\end{align}

In the limit of negligible SPM of the pump $\theta(z,q)\ll1$, the
pump spectral function $\beta_{p}(z,\omega)$ becomes independent
of $z$ and the right hand side of equations of motion (\ref{eomszw})
depends only on $z$ via the prefactors $\gamma_{\text{SPM},s}$,
$\gamma_{\text{SPM},i}$, $\gamma_{\delta}$. 
However, as soon as SPM becomes important this simple translational
dependence is lost. Also note that the SPDC pump spectral function
($\delta=1$) in Eq. (\ref{pumpspect}), satisfies 
\begin{align}
\int d\omega|\beta_{p}(z,\omega)|^{2}&=\mathcal{E}_p(0)\\
&=\hbar\bar{\omega}_{p}\int dz|\Lambda(z)|^{2}=\hbar\bar{\omega}_{p}N_{p}=E_{p}, \nonumber
\end{align}
where $E_p$ is the energy contained in the pump pulse. 
Having introduced the operators $a_{j}(z,\omega)$ and their equations
of motion, we would like to study their equal $z$ commutation relation.
For example, 
\begin{align}
[a_{j}(z,\omega),a_{j}^{\dagger}(z,\omega')]=\frac{v_{j}}{2\pi}\int & dtdt'e^{i\omega(t-z/v_{p})-i\omega'(t'-z/v_{p})}\nonumber \\
 & \times[\psi_{j}(z,t),\psi_{j}^{\dagger}(z,t')],
\end{align}
which shows that to know the equal position commutator of the $a_{j}(z,\omega)$
it is necessary to know the \emph{unequal} time commutator $[\psi_{j}(z,t),\psi_{j}^{\dagger}(z,t')]$.
To know this commutator requires, in principle, knowledge of  the
dynamics  of the field operators $\psi_{j}(z,t)$ for all times, as
given in Eq. (\ref{eomszt}). Despite this difficulty, partial progress
can be made for positions $z_{m}=z_{0}<\ell_{\min}$ or $z_{m}=z_{1}>\ell_{\max}$
before or after the nonlinear region, where one can use the following
identity 
\begin{align}
\psi_{j}(z_{m},t)=e^{-i\bar{\omega}_{j}(t-t_{m})}\psi_{j}(z_{m}-v_{j}(t-t_{m}),t_{m}),\label{fp}
\end{align}
where  $t_{m}=t_{0}$ or $t_{m}=t_{1}$ are times chosen respectively
before and after there is any nonlinear coupling, to show that 
\begin{align}
[{\psi}_{j} & (z_{m},t),\psi_{j}^{\dagger}(z_{m},t')]\\
 & =[\psi_{j}(z_{m}-v_{j}(t-t_{m}),t_{m}),\psi_{j}^{\dagger}(z_{m}-v_{j}(t'-t_{m}),t_{m})]\nonumber \\
 & =\delta(v_{j}(t-t')),\nonumber 
\end{align}
and use that result to show that for positions \emph{outside} the
nonlinear region \begin{subequations}\label{CCRzw} 
\begin{align}
[a_{j}(z_{m},\omega),a_{j'}^{\dagger}(z_{m},\omega')] & =\delta_{j,j'}\delta(\omega-\omega'),\\{}
[a_{j}(z_{m},\omega),a_{j'}(z_{m},\omega')] & =0.
\end{align}
\end{subequations} In the next section we will come back to this
question and show that, indeed, the commutation relations Eq. (\ref{CCRzw}a,\ref{CCRzw}b)
hold for all $z$, both inside and outside the nonlinear region. These
allows us to interpret quantities such as 
\begin{align}
a_{j}^{\dagger}(z,\omega)a_{j}(z,\omega)
\end{align}
as a photon frequency density at position $z$, in such a way that
the total number of photons passing through a plane cutting the waveguide
at $z$ is precisely $\int d\omega \  a_{j}^{\dagger}(z,\omega)a_{j}(z,\omega)$.

\section{Solving the equations of motion}

\label{soleoms} For computational purposes and notational simplicity
we discretize the operators $a_{j}(z,\omega)$ on a grid of $N$ points
according to $\omega_{n}=\omega_{0}+n\Delta\omega|_{n=0}^{N-1}$.
We introduce the column vectors $\mathbf{u}$ and $\mathbf{v}^{\dagger}$
with components 
\begin{align}
\mathbf{u}_{n}(z)=a_{s}(z,\omega_{n}),\\
\mathbf{v}_{n}^{\dagger}(z)=a_{i}^{\dagger}(z,\omega_{n}),
\end{align}
and then, using Eq. (\ref{eomszw}) we can write 
\begin{align}
\frac{\partial}{\partial z}\left(\begin{array}{c}
\mathbf{u}(z)\\
\mathbf{v}^{\dagger}(z)
\end{array}\right)=i\underbrace{\left[\begin{array}{c|c}
\mathbf{G}(z) & \mathbf{F}(z)\\
\hline -\mathbf{F}^{\dagger}(z) & -\mathbf{H}^{\dagger}(z)
\end{array}\right]}_{:=\mathbf{Q}(z)}\left(\begin{array}{c}
\mathbf{u}(z)\\
\mathbf{v}^{\dagger}(z)
\end{array}\right),\label{qdef}
\end{align}
where we have defined the following matrices \begin{subequations}
\begin{align}
\mathbf{F}_{n,m}(z) & =\frac{\gamma_{\delta}g(z)}{\sqrt{2\pi}}\beta_{p}(z,\omega_{n}+\omega_{m})\Delta\omega,\\
\mathbf{G}_{n,m}(z) & =\Delta k_{s}(\omega_{n})\delta_{m,n}+\frac{\gamma_{\text{XPM},s}h_{s}(z)}{2\pi}\mathcal{E}_p(\omega_{n}-\omega_{m})\Delta\omega,\\
\mathbf{H}_{n,m}(z) & =\Delta k_{i}(\omega_{n})\delta_{m,n}+\frac{\gamma_{\text{XPM},i}h_{i}(z)}{2\pi}\mathcal{E}_p^{*}(\omega_{n}-\omega_{m})\Delta\omega.
\end{align}
\end{subequations} We can now formally integrate the discretized
equations of motion and obtain 
\begin{align}
\left(\begin{array}{c}
\mathbf{u}(z)\\
\mathbf{v}^{\dagger}(z)
\end{array}\right) & =\mathbf{U}(z,z_{0})\left(\begin{array}{c}
\mathbf{u}(z_{0})\\
\mathbf{v}^{\dagger}(z_{0})
\end{array}\right)\label{solK}\\
 & =\left[\begin{array}{c|c}
\mathbf{U}^{s,s}(z,z_{0}) & \mathbf{U}^{s,i}(z,z_{0})\\
\hline (\mathbf{U}^{i,s}(z,z_{0}))^{*} & (\mathbf{U}^{i,i}(z,z_{0}))^{*}
\end{array}\right]\left(\begin{array}{c}
\mathbf{u}(z_{0})\\
\mathbf{v}^{\dagger}(z_{0})
\end{array}\right),
\end{align}
where the propagator $\mathbf{U}(z,z_{0})$ is defined by the limit
\begin{align}
\mathbf{U}(z,z_{0})=\lim_{n\to\infty}\prod_{p=1}^{n}\exp\left(i\Delta z\mathbf{Q}(z_{p})\right),
\end{align}
and $\Delta z=(z-z_{0})/n$ and $z_{p}=z_{0}+p\Delta z$. The intuition
behind the Trotter-Suzuki expansion used in the last equation is that
for sufficiently thin ``slices'' of propagation in $z$ one can
approximate the matrix $\mathbf{Q}(z)$ as a constant in that region;
thus one can simply compound the propagation over all the small regions
to get the net propagator. Finally, note that if $\mathbf{Q}$ is
independent of $z$  then 
\begin{align}
\mathbf{U}(z,z_{0}) & =\lim_{n\to\infty}\prod_{p=1}^{n}\exp\left(i\Delta z\mathbf{Q}\right)=\exp(i(z-z_{0})\mathbf{Q}).\label{zind}
\end{align}
This will always be the case for a uniform waveguide in the limit where the SPM of the pump is negligible. 

The undiscretized form of Eq. (\ref{solK}) yields the linear transformation
of the continuous-frequency $(z,\omega)$ operators \begin{subequations}\label{heissol}
\begin{align}
a_{s}(z,\omega)= & \int d\omega'U^{s,s}(\omega,\omega';z,z_{0})a_{s}(z_{0},\omega')\\
 & +\int d\omega'U^{s,i}(\omega,\omega';z,z_{0})a_{i}^{\dagger}(z_{0},\omega'),\nonumber \\
a_{i}^{\dagger}(z,\omega)= & \int d\omega'(U^{i,s}(\omega,\omega';z,z_{0}))^{*}a_{s}(z_{0},\omega')\\
 & +\int d\omega'(U^{i,i}(\omega,\omega';z,z_{0}))^{*}a_{i}^{\dagger}(z_{0},\omega'),\nonumber 
\end{align}
\end{subequations} where the blocks of the propagator $\mathbf{U}(z,z_{0})$
are related to the continuous-frequency transfer functions as follows:
\begin{align}
U^{j,k}(\omega_{n},\omega_{m};z,z_{0}) & =\mathbf{U}_{n,m}^{j,k}(z,z_{0})/\Delta\omega.\label{mattotrans}
\end{align}

For notational simplicity, we omit the spatial dependence when we
write the transfer functions connecting the input and output operators
in the distant past and future. Defining $U^{j,j}(\omega,\omega')=U^{j,j}(\omega,\omega',z_1,z_0) e^{i \Delta k_j(\omega') z_0 -i \Delta k_j(\omega) z_1}$,
$U^{j,l}(\omega,\omega')=U^{j,l}(\omega,\omega',z_1,z_0) e^{-i \Delta k_j(\omega') z_0 -i \Delta k_l(\omega) z_1} $ ($j \neq l$)
as well as $a_{l}^{\text{(in/out)}}(\omega)=
e^{-i \Delta k_l(\omega) z_{0/1}}
a_{l}(z_{0/1},\omega)$
we write

\begin{subequations} 
\begin{align}
a_{s}^{\text{(out)}}(\omega)= & \int d\omega'U^{s,s}(\omega,\omega')\ a_{s}^{\text{(in)}}(\omega')\\
 & +\int d\omega'U^{s,i}(\omega,\omega')\ a_{i}^{\dagger\text{(in)}}(\omega'),\nonumber \\
a_{i}^{\text{(out)}}(\omega)= & \int d\omega'U^{i,i}(\omega,\omega')\ a_{i}^{\text{(in)}}(\omega')\\
& +\int d\omega'U^{i,s}(\omega,\omega')\ a_{s}^{\dagger\text{(in)}}(\omega').\nonumber 
\end{align}
\end{subequations} 

\
 The propagator $\mathbf{U}(z,z_{0})$ allows us to write the operators
in the spatial region after the nonlinear region, $a_{j}(z_{1},\omega)$
and $a_{j}^{\dagger}(z_{1},\omega)$, as linear combinations of the
operators before the nonlinear region $a_{j}(z_{0},\omega')$ and
$a_{j}^{\dagger}(z_{0},\omega')$. This is not, however, a solution
of Heisenberg's equations; the latter, as in Eq. (\ref{haction})
would allow us to write \emph{time evolving} operators in the distant
future in terms of the operators in the distant past. However, using
the results from Appendix \ref{solzt}, one can show that \begin{subequations}\label{inout}
\begin{align}
\left.b_{jk}(t_{0})\right|_{k=\bar{k}_{j}+(\omega-\bar{\omega_{j}})/v_{j}} & =
\sqrt{v_{j}}e^{-i\omega t_{0}-i\Delta k_{j}(\omega)z_{0}}a_{j}(z_{0},\omega),\\
\left.b_{jk}(t_{1})\right|_{k=\bar{k}_{j}+(\omega-\bar{\omega_{j}})/v_{j}} & =\sqrt{v_{j}}e^{-i\omega t_{1}-i\Delta k_{j}(\omega)z_{1}}a_{j}(z_{1},\omega),
\end{align}
\end{subequations} allowing us to link the (proper, evolving-in-time)
Heisenberg operators $b_{jk}(t)$ with the operators $a_{j}(z,\omega)$,
and showing that they are the same operators in the distant past and
future (modulo some phases and constant prefactors). Upon realizing
this identity, it is immediately recognizable that the Heisenberg
equations of motion have been solved, since now we can write the Heisenberg
operators $b_{jk}(t_{1})$ in the future in terms of the Heisenberg
operators $b_{jk}(t_{0})$ in the past. This is easily seen by inverting
the relations in Eqs. (\ref{inout}a,\ref{inout}b) and using them
to replace $a_{j}(z_{0},\omega)$ and $a_{j}(z_{1},\omega)$ by $b_{jk}(t_{0})$
and $b_{jk}(t_{1})$ in the right and left hand sides of Eqs. (\ref{heissol}a,\ref{heissol}b)
with $z=z_{1}$.

\section{Commutation relations and modal structure}

We now go back to the question posed at the end of section \ref{sec:dynamics}
and analyze the equal $z$ commutators of the fields inside the nonlinear
medium which, upon using the solutions in Eq. (\ref{heissol}) and
the initial position commutators in Eq. (\ref{CCRzw}), we find to
be \begin{subequations}\label{putative1} 
\begin{align}
[a_{s}(z,\omega), & a_{s}^{\dagger}(z,\omega')]=\\
 & \int d\omega''U^{s,s}(\omega,\omega'';z,z_{0})(U^{s,s}(\omega',\omega'';z,z_{0}))^{*}\nonumber \\
 & -\int d\omega''U^{s,i}(\omega,\omega'';z,z_{0})(U^{s,i}(\omega',\omega'';z,z_{0}))^{*},\nonumber \\{}
[a_{i}(z,\omega), & a_{i}^{\dagger}(z,\omega')]=\\
 & \int d\omega''U^{i,i}(\omega,\omega'';z,z_{0})(U^{i,i}(\omega',\omega'';z,z_{0}))^{*}\nonumber \\
 & -\int d\omega''U^{i,s}(\omega,\omega'';z,z_{0})(U^{i,s}(\omega',\omega'';z,z_{0}))^{*},\nonumber \\{}
[a_{s}(z,\omega), & a_{i}(z,\omega')]=\\
 & \int d\omega''U^{s,s}(\omega,\omega'';z,z_{0})U^{i,s}(\omega',\omega'';z,z_{0})\nonumber \\
 & -\int d\omega''U^{s,i}(\omega,\omega'';z,z_{0})U^{i,i}(\omega',\omega'';z,z_{0}),\nonumber 
\end{align}
\end{subequations} and $[a_{s}(z,\omega),a_{i}^{\dagger}(z,\omega')]=0$.
To show that the right hand sides of Eqs. (\ref{putative1} a,b,c)
are $\delta(\omega-\omega')$, $\delta(\omega-\omega')$, and $0$ respectively,
we note that the matrix discretized versions of these putative commutations
relations would be \begin{subequations} 
\begin{align}
\mathbf{U}^{s,s}(z,z_{0})(\mathbf{U}^{s,s}(z,z_{0}))^{\dagger}-\mathbf{U}^{s,i}(z,z_{0})(\mathbf{U}^{s,i}(z,z_{0}))^{\dagger} & =\mathbb{I}_{N},\\
\mathbf{U}^{i,i}(z,z_{0})(\mathbf{U}^{i,i}(z,z_{0}))^{\dagger}-\mathbf{U}^{i,s}(z,z_{0})(\mathbf{U}^{i,s}(z,z_{0}))^{\dagger} & =\mathbb{I}_{N},\\
\mathbf{U}^{s,s}(z,z_{0})(\mathbf{U}^{i,s}(z,z_{0}))^{T}-\mathbf{U}^{s,i}(z,z_{0})(\mathbf{U}^{i,i}(z,z_{0}))^{T} & =0,
\end{align}
\end{subequations} with $\mathbb{I}_{N}$ being the $N$ dimensional
identity matrix. Note that the last set of equations can be written
more compactly in terms of the following equation for the propagator
$\mathbf{U}(z,z_{0})$ 
\begin{align}
\mathbf{U}(z,z_{0})\ \mathbf{S}\ \mathbf{U}^{\dagger}(z,z_{0})=\mathbf{S},\label{SU11}
\end{align}
with 
\begin{align}
\mathbf{S}=\begin{bmatrix}\mathbb{I}_{N} & 0\\
0 & -\mathbb{I}_{N}
\end{bmatrix}.
\end{align}
Mathematically, Eq. (\ref{SU11}) states that $\mathbf{U}(z,z_{0})$
is an element of the $SU(1,1)$ Lie group (cf. Appendix 11.1.4. of
Klimov and Chumakov \cite{klimov2009group}). To show that $\mathbf{U}(z,z_{0})\in SU(1,1)$
it is sufficient to show that its generators, the matrices $\mathbf{Q}(z)$
belong to the algebra of this group, $\mathfrak{su}(1,1)$, thus they
need to satisfy 
\begin{align}
\mathbf{Q}(z)\ \mathbf{S}=\mathbf{S}\ \mathbf{Q}^{\dagger}(z).
\end{align}
But this is trivial to show using the Hermiticity of the matrices
$\mathbf{G}$ and $\mathbf{H}$ that, together with $\mathbf{F}$,
define $\mathbf{Q}$ in Eq. (\ref{qdef}). Thus, the bonafide commutation
relations of the $a_{j}(z,\omega)$ are guaranteed by the algebraic
structure of the equations of motion it satisfies, together with the
initial conditions for the commutators derived (Eq. (\ref{CCRzw})).
Because of the Lie group constraints, the transfer functions can be
\emph{jointly} decomposed as follows \begin{subequations}\label{expansion1}
\begin{align}
U^{s,s}(\omega,\omega';z,z_{0}) & =\sum_{l}\text{cosh}(r_{l})[\rho_{s}^{(l)}(\omega)][\tau_{s}^{(l)}(\omega')]^{*},\\
U^{s,i}(\omega,\omega';z,z_{0}) & =\sum_{l}\text{sinh}(r_{l})[\rho_{s}^{(l)}(\omega)][\tau_{i}^{(l)}(\omega')],\\
(U^{i,i}(\omega,\omega';z,z_{0}))^{*} & =\sum_{l}\text{cosh}(r_{l})[\rho_{i}^{(l)}(\omega)]^{*}[\tau_{i}^{(l)}(\omega')],\\
(U^{i,s}(\omega,\omega';z,z_{0}))^{*} & =\sum_{l}\text{sinh}(r_{l})[\rho_{i}^{(l)}(\omega)]^{*}[\tau_{s}^{(l)}(\omega')]^{*},
\end{align}
\end{subequations} where the quantities $r_{l}$ are the squeezing
parameter of the Schmidt mode $l$ and the sets of functions $\left\{ \rho_{s,i}^{(l)}\right\} $,
$\left\{ \tau_{s,i}^{(l)}\right\} $ are complete and orthonormal,
and thus for example \begin{subequations}\label{ortho1} 
\begin{align}
\int d\omega\ \rho_{s}^{(l)}(\omega)\ [\rho_{s}^{(l')}(\omega)]^{*} & =\delta_{l,l'},\label{complete}\\
\sum_{l}\rho_{s}^{(l)}(\omega)\ [\rho_{s}^{(l)}(\omega')]^{*} & =\delta(\omega-\omega').
\end{align}
\end{subequations}

\section{Solving the spontaneous problem}

\label{sec:example}

Given the linearity of the input-output relations on the operators,
the state generated when these are applied on vacuum must be Gaussian.
In particular,  in the distant future it will have the form
\begin{align}
 & \ket{\text{TMSV}}=\\
 & \ \exp\left(\int d\omega d\omega'J(\omega,\omega')a_{s}^{\text{(in)}\dagger}(\omega)a_{i}^{\text{(in)}\dagger}(\omega')-\text{H.c.}\right)\ket{\text{vac}}.\nonumber 
\end{align}

This squeezed state is described univocally by its first and second
moments. These are easily constructed once the scattering matrix $\mathbf{U}$
is known. For the sake of illustration, the covariance between signal
and idler annihilation operators is 
\begin{align}
M(\omega,\omega') & =\braket{\text{vac}|a_{s}^{\text{(out)}}(\omega)a_{i}^{\text{(out)}}(\omega')|\text{vac}}\\
 & =\int d\omega''U^{i,i}(\omega,\omega'')U^{s,i}(\omega',\omega'')\nonumber \\
 & =\sum_{l}\frac{\sinh(2r_{l})}{2}\ \rho_{s}^{(l)}(\omega)\rho_{i}^{(l)}(\omega'),\nonumber 
\end{align}
where $\ket{\text{vac}}$ is the vacuum state which is annihilated
by the distant past (input) operators 
\begin{align}
a_{j}^{\text{(in)}}(\omega)\ket{\text{vac}}=a_{j}(z_{0},\omega)\ket{\text{vac}}=b_{jk}(t_{0})\ket{\text{vac}}=0.
\end{align}
From the moments $M$, one easily reconstructs the joint spectral
amplitude in terms of the Schmidt modes and squeezing parameters of
the scattering matrix $\mathbf{U}$,
finding 
\begin{align}
J(\omega,\omega')=\sum_{l}r_{l}\ \rho_{s}^{(l)}(\omega)\rho_{i}^{(l)}(\omega').\label{jsadef}
\end{align}
Note that in the low gain regime $r_{l}\ll1$ one can approximate
$\sinh(2r_{l})/2\approx r_{l}$ and thus $M(\omega,\omega')\approx J(\omega,\omega')$,
but in the high gain regime the relation between the two functions
is more complicated

We can use these results to study what is perhaps the simplest case
of twin-beams generation: a $\chi^{(2)}$ process in which the nonlinearity
has a flat top-hat profile and we ignore any effect of cross- and
self-phase modulation. For a use of the theory presented here in the
characterization of PDC sources involving the aforementioned $\chi^{(3)}$ effects, see our companion paper
\cite{triginer2018complete}.

With the modification of the pump function  by SPM neglected
and the nonlinearity function $\xi_{1}(z)$ a top-hat function
extending from $\ell_{\min}=-\tfrac{\ell}{2}$ to $\ell_{\max}=\tfrac{\ell}{2}$, the matrix $\mathbf{Q}$ in Eq. (\ref{qdef}) is independent of $z$ in the region where the nonlinearity is present. Because of this we can write (recall
Eq. (\ref{zind})) 
\begin{align*}
\mathbf{U}\left(-\tfrac{\ell}{2},\tfrac{\ell}{2}\right)=\exp(i\mathbf{Q}\ell),
\end{align*}
and the calculation of the matrix propagator $\mathbf{U}$ is reduced
to a single exponentiation, which is one of the main advantages of working
with the $a(z,\omega)$ operators instead of the $\psi(z,t)$ operators
\cite{vidrighin2017quantum}.

For illustration, we study a Gaussian pump 
\begin{align}
 & \braket{\psi_{p}(z,t_{0})}=
 \frac{\sqrt{N_{p}}}{\sqrt[4]{\pi(v_{p}/\sigma)^{2}}}\exp\left(-\frac{(z-{z}_{0})^{2}}{2(v_{p}/\sigma)^{2}}\right),
\nonumber 
\end{align}
 localized around $z={z}_{0}$ at time $t=t_{0}$, and with bandwidth
$\sigma$ and mean number of photons $N_{p}$. 

The low gain joint spectral amplitude (JSA), in the limit where the spectral content of the pump
is not modified, is simply  
\begin{align}
J(\omega,\omega')
= & \frac{\xi_{1}^{(0)}\sqrt{N_{p}}}{\sqrt{2\pi v_{s}v_{i}v_{p}\sigma\sqrt{\pi}}}\exp\left({-\frac{(\omega+\omega'-\bar{\omega}_{p})^{2}}{2\sigma^{2}}}\right) \nonumber \\
 & \times\ell\ \text{sinc}\left(\tfrac{\ell}{2}\left\{ \Delta k_{s}(\omega)+\Delta k_{i}(\omega')\right\} \right). \label{eq:lowgainJSA}
\end{align}
 \cite{grice1997spectral} (see Appendix \ref{lowgain} for a derivation),
where $\xi_{1}^{(0)}$ is the nonzero value the nonlinearity function
$\xi_{1}(z)$ takes in the the region $-\ell/2<z<\ell/2$.

We will work in the symmetric group velocity matched regime \cite{graffitti2018design}, where $(v_p^{-1} - v_s^{-1}) = -(v_p^{-1} - v_s^{-1}) = 2\kappa/\ell $, to obtain
\begin{align}
\tfrac{\ell}{2}\left\{ \Delta k_{s}(\omega)+\Delta k_{i}(\omega')\right\} =\kappa\left\{ (\omega-\bar{\omega}_{s})-(\omega'-\bar{\omega}_{i})\right\} ,
\end{align}
and furthermore pick the parameter $\kappa=1.61/(1.13\sigma)$ so
as to maximize the separability of the low gain JSA in Eq. (\ref{eq:lowgainJSA})
by matching the width of the sinc function and the Gaussian appearing
appearing there \cite{graffitti2018design}. 
\begin{figure}
\includegraphics[width=0.95\linewidth]{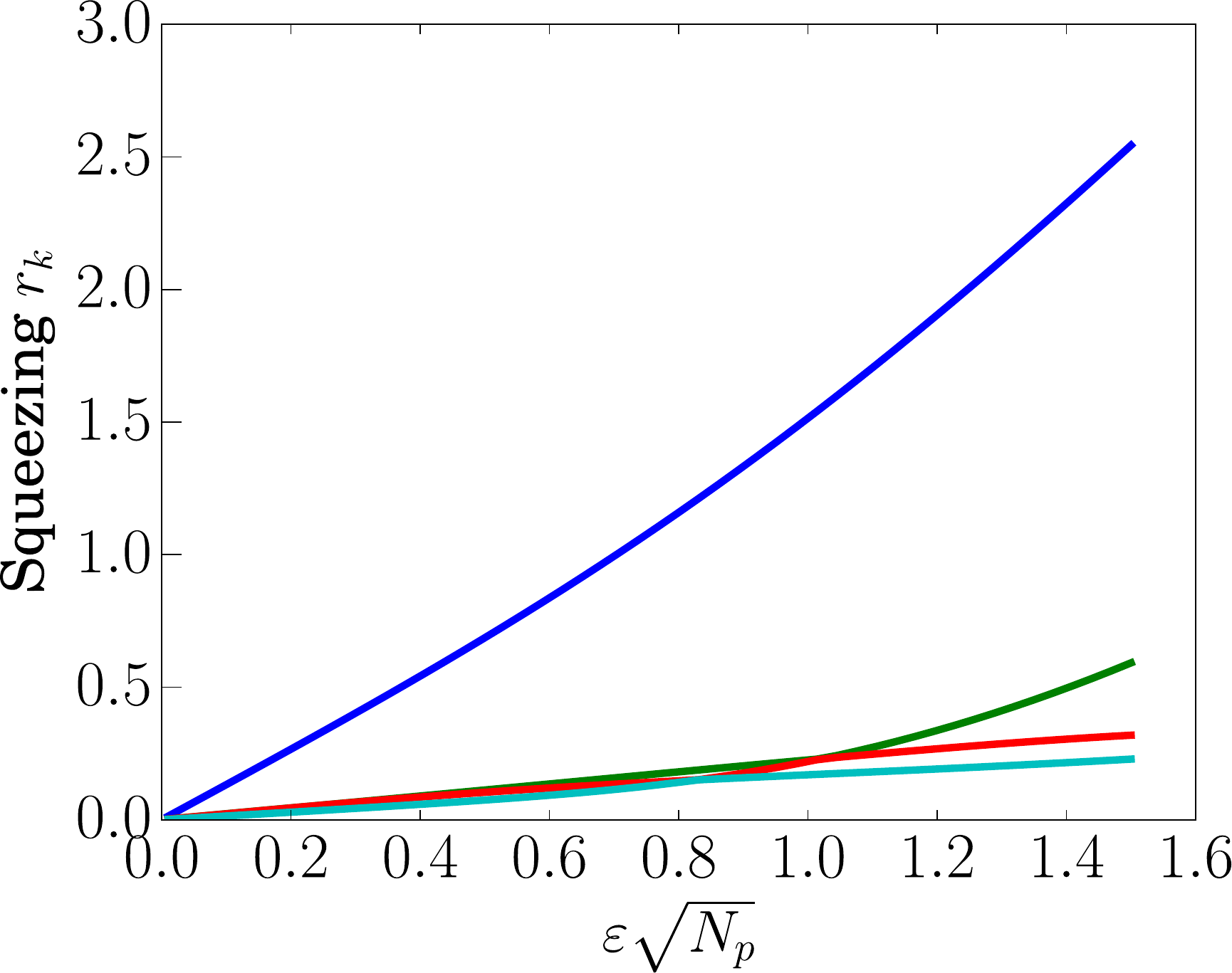}\
 \caption{We plot the squeezing parameters $\{r_{k}\}$ of the four largest Schmidt
modes. For low gain the squeezing parameters $r_{k}$ are linear
in $\sqrt{N_{p}}$.  However, in the region where $\varepsilon \sqrt{N_p} \gtrsim 1$ the dependence of the two largest squeezing parameters on that variable deviates from linear.}
\label{fig:squeezing} 
\end{figure}

\begin{figure}
\includegraphics[width=0.47\linewidth]{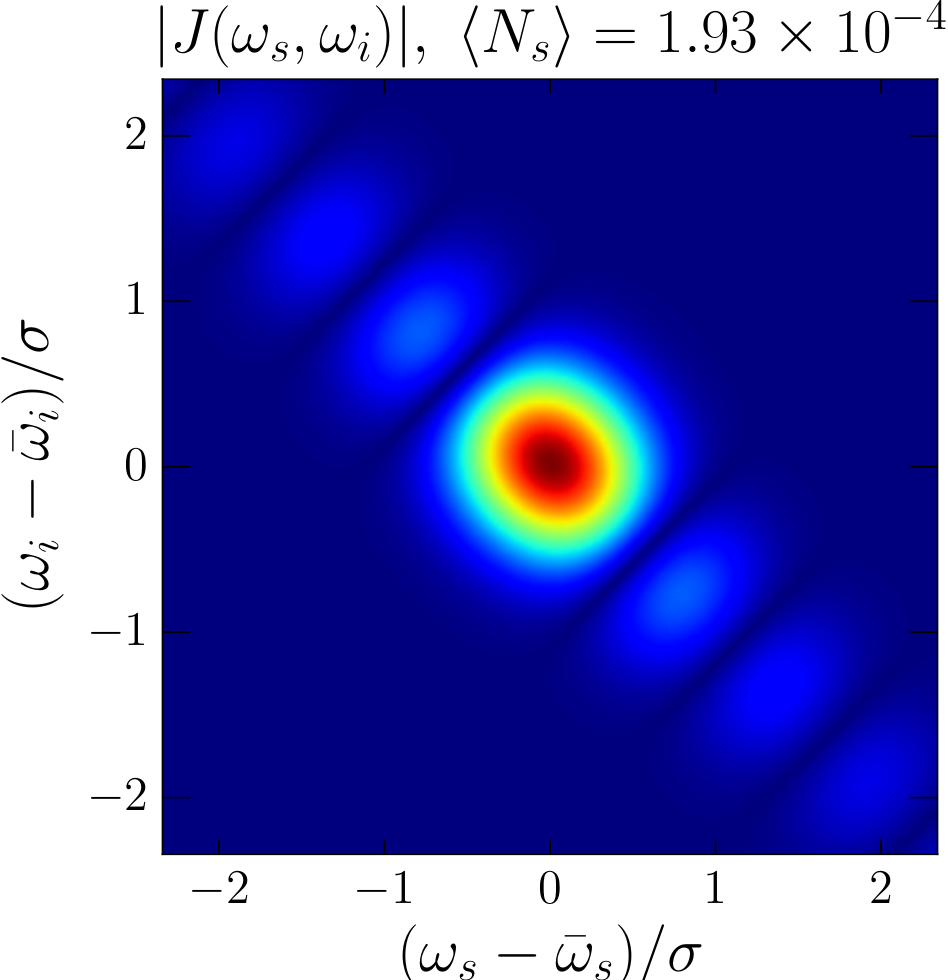}\
 \includegraphics[width=0.47\linewidth]{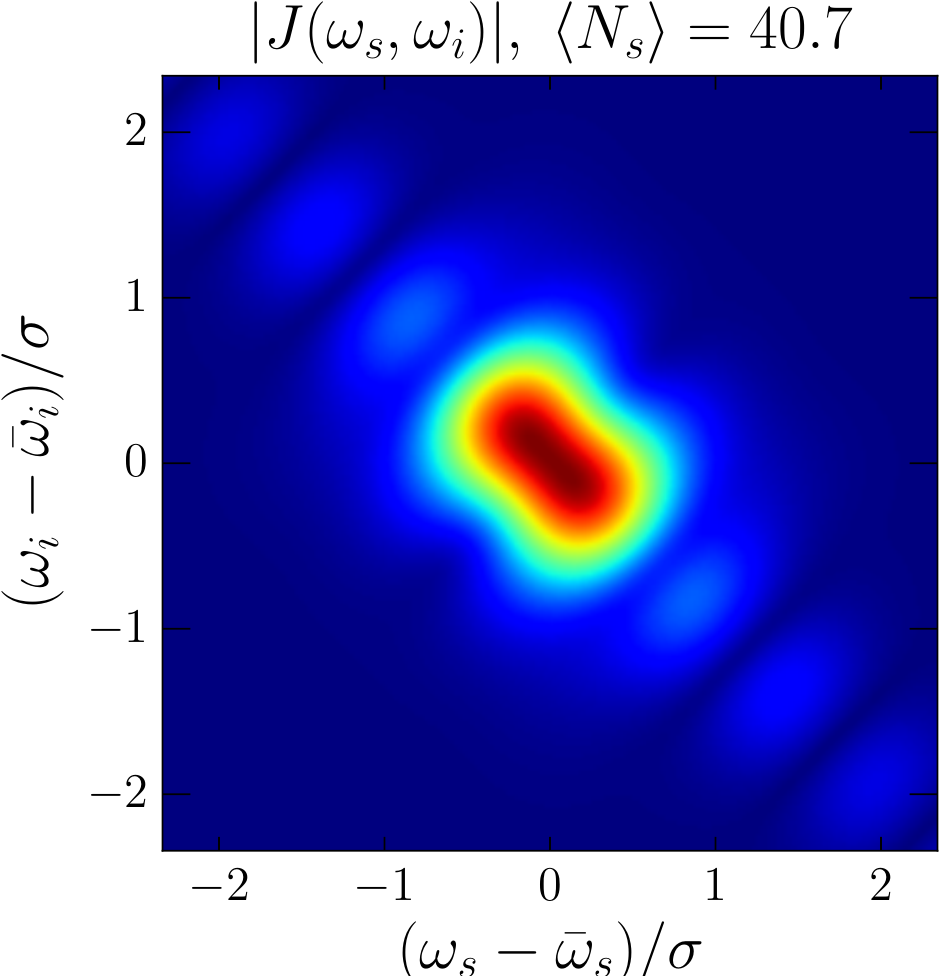}\
 \caption{Absolute values of the joint spectral amplitude in the symmetric group
velocity matched regime in the low (left panel) and high (right panel)
gain regime. In the low-gain regime the JSA is simply the product
of the pump function (Gaussian) and the phase-matching function (sinc).
In the high-gain regime this is not the case because of so-called
time-ordering corrections \cite{quesada2014effects,quesada2015time}.}
\label{fig:jsas} 
\end{figure}

In Fig. \ref{fig:squeezing} we show the evolution of the squeezing
parameters of the JSA from the low-gain regime to the high-gain regime
as the pump energy $N_{p}$ is increased. As predicted using the
Magnus expansion \cite{quesada2014effects,quesada2015time}, the time-ordering corrections will cause the squeezing parameters to behave in a nonlinear way as a function of $\sqrt{N_{p}}$.
Note that this result will also be observed regardless of the shape
of the pump function and the profile of the nonlinearity. In particular
these time-ordering corrections will also affect the optimal Gaussian
PMF/Gaussian pump function combination that  uniquely gives a fully
factorable JSA in the low gain regime \cite{Uren2005,quesada2018gaussian}.

In Fig. \ref{fig:jsas} we also show the JSA as defined in Eq. (\ref{jsadef})
in the low-gain regime $\braket{N_{s}}=\braket{N_{i}}\ll1$ and in
the high-gain regime where the mean number of photons in the signal
and idler beams is $\braket{N_{s}}=\braket{N_{i}}=41$ with 
\begin{align}
\braket{N_{j}}=\int d\omega\braket{a_{j}^{\text{(out)}\dagger}(\omega)a_{j}^{\text{(out)}}(\omega)}=\sum_{l}\sinh^{2}(r_{l}).
\end{align}

The computation times for each JSA for a given value of $N_{p}$ and
for a grid of 600 frequencies takes seconds on a desktop computer
using \texttt{Python}'s\cite{oliphant2007python} \texttt{scipy}\cite{scipy};
this time should be contrasted with the hours it takes with other
methods and publicly available code \cite{christ2013theory,christ2013code}
running in the same hardware and language/libraries.

\section{Conclusions and outlook}

\label{sec:conc} We have presented a justification for the use of
field operators in $(z,\omega)$ space in the study of twin-beam generation.
These operators have been constructed rigorously, starting from a
canonical formalism that has Maxwell's equations as the Heisenberg
equations of motion. In the limit of negligible group velocity dispersion,
we showed that the $a_{j}(z,\omega)$ operators satisfy well-defined
equal position commutator relations. Furthermore, we showed that
for times and positions long before/after the pump has entered/exited
the nonlinear region, these position-evolving operators indeed coincide
with standard Heisenberg operators evolving in the standard Heisenberg
picture in time. The solution to the equations these operators satisfy
is easy to implement computationally, and allows for the incorporation
of many important processes that can alter the properties of the twin-beams,
such as poling inhomogeneities (via $\xi_{1}(z)$), self-phase modulation
of the pump, and cross-phase modulation of the pump on the twin-beams.
A thorough exploration of this m\'elange of wave mixings is presented in our companion
paper \cite{triginer2018complete}. 

The derivation presented for the $(z,\omega)$ operators is apparently   not easily generalizable
to include the presence of   group velocity dispersion. Intuitively,
if dispersion is important, position is not like time and wavevectors
are not the same as frequencies. Mathematically, if the relation between
frequencies and wavevectors is nonlinear, then one cannot obtain identities
such as Eq. (\ref{fp}), and thus one cannot (at least in an obvious
manner) prove the bona fide commutation relation of the $a_{j}(z,\omega)$
operators at equal positions. Indeed, we show in Appendix \ref{app:comms}
that certain commutators of the $z,\omega$ operators that should
be trivially zero are non zero when dispersion is included. 

In principle, one can take dispersion into account by solving the dynamics of the Heisenberg operators that evolve \emph{in time} by generalizing Eq.~\eqref{eomszt} to include a nonlinear dispersion relation \cite{helt2019degenerate}. Yet, for many applications in quantum nonlinear optics, it is sufficient and often necessary to work with narrow enough bandwidths, such that group velocities are well defined. This is especially true when generating twin-beams with a small degree of frequency correlations \cite{quesada2018gaussian}.

Finally, we would like to point out that the methods presented here
can easily be carried over to frequency conversion, where now the
fields $a_{s}(z,\omega)$ will couple to $a_{i}(z,\omega')$ instead
of $a_{i}^{\dagger}(z,\omega')$. In this case, the underlying group
dictating the symmetry of the problem will be $SU(2)$. The generalization
of the techniques presented here should provide a useful tool to study
highly mode-selective frequency conversion beyond the perturbative
regime\cite{christ2013theory,reddy2014efficient,christensen2015temporal,quesada2016high}.

\section*{Acknowledgement}

N.Q. and J.E.S. thank the National Science and Engineering Research
Council of Canada. G.T. thanks Merton College, Oxford, for its support.
M.D.V. thanks the Engineering and Physical Sciences Research Council
for funding through grant EP/K034480/1 (BLOQS).

\emph{Note added ---} While preparing this manuscript we became aware of related work by Sharapova et al. \cite{sharapova2020properties} where equations similar to the ones derived here are used to study the joint spectral amplitude of the transverse degrees of freedom of a bright squeezing source.

\bibliographystyle{unsrt}
 \bibliography{review.bib}

\onecolumngrid

\appendix

\section{Linear Hamiltonian in terms of the field operators}

\label{app:linearpsi} Expanding $\omega_{\mu k}$ about $\omega_{\mu k_{\mu}}\equiv\bar{\omega}_{\mu}$
we can write the Hamiltonian (\ref{eq:DandB-1}) as 
\begin{align*}
H_{L}= & \sum_{\mu}\hbar\bar{\omega}_{\mu}\;\int dk\;b_{\mu k}^{\dagger}b_{\mu k}+\sum_{\mu}\hbar v_{\mu}\;\int dk\;(k-k_{\mu})b_{\mu k}^{\dagger}b_{\mu k}+\frac{1}{2}\sum_{\mu}\hbar v_{\mu}^{\prime}\;\int dk\;(k-k_{\mu})^{2}b_{\mu k}^{\dagger}b_{\mu k}+\cdots,
\end{align*}
where 
\begin{align}
v_{\mu}= & \left(\frac{d\omega_{\mu k}}{dk}\right)_{k_{\mu}},\quad v_{\mu}^{\prime}=\left(\frac{d^{2}\omega_{\mu k}}{dk^{2}}\right)_{k_{\mu}}.\label{eq:GV=000026GVD}
\end{align}
Since from (\ref{psidef}) we can write 
\begin{align}
b_{\mu k}^{\dagger}b_{\mu k}=\int\frac{dzdz^{\prime}}{2\pi}\psi_{\mu}^{\dagger}(z)\psi_{\mu}(z^{\prime})e^{i(k-k_{\mu})(z-z^{\prime})},
\end{align}
we have 
\begin{align}
\int dk\;b_{\mu k}^{\dagger}b_{\mu k}=\int dz\;\psi_{\mu}^{\dagger}(z)\psi_{\mu}(z),
\end{align}
while
\begin{align}
\int dk\;(k-k_{\mu})b_{\mu k}^{\dagger}b_{\mu k}= & \frac{1}{2i}\int\frac{dkdzdz^{\prime}}{2\pi}\psi_{\mu}^{\dagger}(z)\psi_{\mu}(z^{\prime})\left[\left(\frac{\partial}{\partial z}-\frac{\partial}{\partial z^{\prime}}\right)e^{i(k-k_{\mu})(z-z^{\prime})}\right]\\
= & \frac{i}{2}\int\frac{dkdzdz^{\prime}}{2\pi}\left[\left(\frac{\partial}{\partial z}-\frac{\partial}{\partial z^{\prime}}\right)\psi_{\mu}^{\dagger}(z)\psi_{\mu}(z^{\prime})\right]e^{i(k-k_{\mu})(z-z^{\prime})}\\
= & \frac{i}{2}\int dz\left(\frac{\partial\psi_{\mu}^{\dagger}(z)}{\partial z}\psi_{\mu}(z)-\psi_{\mu}^{\dagger}(z)\frac{\partial\psi_{\mu}(z)}{\partial z}\right),
\end{align}
and 
\begin{align}
\int dk\;(k-k_{\mu})^{2}b_{\mu k}^{\dagger}b_{\mu k}= & \int\frac{dkdzdz^{\prime}}{2\pi}\psi_{\mu}^{\dagger}(z)\psi_{\mu}(z^{\prime})\left[\left(\frac{\partial}{\partial z}\frac{\partial}{\partial z^{\prime}}\right)e^{i(k-k_{\mu})(z-z^{\prime})}\right]\\
= & \int\frac{dkdzdz^{\prime}}{2\pi}\left[\left(\frac{\partial}{\partial z}\frac{\partial}{\partial z^{\prime}}\right)\psi_{\mu}^{\dagger}(z)\psi_{\mu}(z^{\prime})\right]e^{i(k-k_{\mu})(z-z^{\prime})}\\
= & \int dz\frac{\partial\psi_{\mu}^{\dagger}(z)}{\partial z}\frac{\partial\psi_{\mu}(z)}{\partial z}.
\end{align}
So the Hamiltonian (\ref{eq:DandB-1}) is
\begin{align}
H_{L}= & \sum_{\mu}\hbar\bar{\omega}_{\mu}\int dz\;\psi_{\mu}^{\dagger}(z)\psi_{\mu}(z)+\frac{i}{2}\sum_{\mu}\hbar v_{\mu}\int dz\left(\frac{\partial\psi_{\mu}^{\dagger}(z)}{\partial z}\psi_{\mu}(z)-\psi_{\mu}^{\dagger}(z)\frac{\partial\psi_{\mu}(z)}{\partial z}\right)\nonumber \\
 & \quad+\frac{1}{2}\sum_{\mu}\hbar v_{\mu}^{\prime}\int dz\frac{\partial\psi_{\mu}^{\dagger}(z)}{\partial z}\frac{\partial\psi_{\mu}(z)}{\partial z}+\cdots
\end{align}

\section{The nonlinear coefficients}

\label{coeff:app} The nonlinear coefficients describing the nonlinear
interaction between the pump, signal and idler modes are defined as
follows \begin{subequations} 
\begin{align}
\zeta_{p}(z) & =\frac{3}{\epsilon_{0}\hbar}\left(\frac{\hbar\bar{\omega}_{p}}{2}\right)^{2}\int dxdy\ \Gamma_{3}^{ijlm}(\mathbf{r})\left(d_{p\bar{k}_{p}}^{i}(x,y)\right)^{*}\left(d_{p\bar{k}_{p}}^{j}(x,y)\right)^{*}d_{p\bar{k}_{p}}^{l}(x,y)d_{p\bar{k}_{p}}^{m}(x,y)\\
 & =\frac{3}{\epsilon_{0}\hbar}\left(\frac{\hbar\bar{\omega}_{p}}{2}\right)^{2}\int dxdy\ \frac{\chi_{3}^{ijlm}(\mathbf{r})}{\epsilon_{0}n_{i}^{2}n_{j}^{2}n_{l}^{2}n_{m}^{2}}\left(d_{p\bar{k}_{p}}^{i}(x,y)\right)^{*}\left(d_{p\bar{k}_{p}}^{j}(x,y)\right)^{*}d_{p\bar{k}_{p}}^{l}(x,y)d_{p\bar{k}_{p}}^{m}(x,y),\\
\zeta_{i/s}(z) & =2\frac{3}{\epsilon_{0}\hbar}\left(\frac{\hbar\bar{\omega}_{{i/s}}}{2}\right)\left(\frac{\hbar\bar{\omega}_{p}}{2}\right)\int dxdy\ \Gamma_{3}^{ijlm}(\mathbf{r})\left(d_{p\bar{k}_{p}}^{i}(x,y)\right)^{*}\left(d_{i,s\bar{k}_{{i,s}}}^{j}(x,y)\right)^{*}d_{i,s\bar{k}_{i,s}}^{l}(x,y)d_{p\bar{k}_{p}}^{m}(x,y)\\
 & =2\frac{3}{\epsilon_{0}\hbar}\left(\frac{\hbar\bar{\omega}_{{i/s}}}{2}\right)\left(\frac{\hbar\bar{\omega}_{p}}{2}\right)\int dxdy\ \frac{\chi_{3}^{ijlm}(\mathbf{r})}{\epsilon_{0}n_{i}^{2}n_{j}^{2}n_{l}^{2}n_{m}^{2}}\left(d_{p\bar{k}_{p}}^{i}(x,y)\right)^{*}\left(d_{i,s\bar{k}_{{i,s}}}^{j}(x,y)\right)^{*}d_{i,s\bar{k}_{i,s}}^{l}(x,y)d_{p\bar{k}_{p}}^{m}(x,y),\\
\xi_{2}(z) & =\frac{3}{\epsilon_{0}\hbar}\left(\frac{\hbar\sqrt{\bar{\omega}_{s}\bar{\omega}_{i}}}{2}\right)\left(\frac{\hbar\bar{\omega}_{p}}{2}\right)\int dxdy\ \Gamma_{3}^{ijlm}(\mathbf{r})\left(d_{s\bar{k}_{s}}^{i}(x,y)\right)^{*}\left(d_{i\bar{k}_{i}}^{j}(x,y)\right)^{*}d_{p\bar{k}_{p}}^{l}(x,y)d_{p\bar{k}_{p}}^{m}(x,y)\\
 & =\frac{3}{\epsilon_{0}\hbar}\left(\frac{\hbar\sqrt{\bar{\omega}_{s}\bar{\omega}_{i}}}{2}\right)\left(\frac{\hbar\bar{\omega}_{p}}{2}\right)\int dxdy\ \frac{\chi_{3}^{ijlm}(\mathbf{r})}{\epsilon_{0}n_{i}^{2}n_{j}^{2}n_{l}^{2}n_{m}^{2}}\left(d_{s\bar{k}_{s}}^{i}(x,y)\right)^{*}\left(d_{i\bar{k}_{i}}^{j}(x,y)\right)^{*}d_{p\bar{k}_{p}}^{l}(x,y)d_{p\bar{k}_{p}}^{m}(x,y),\\
\xi_{1}(z) & =\frac{2}{\epsilon_{0}\hbar}\sqrt{\frac{\hbar^{3}\bar{\omega}_{i}\bar{\omega}_{s}\bar{\omega}_{p}}{(2)^{3}}}\int dxdy\ \Gamma_{2}^{ijl}(\mathbf{r})\left(d_{i\bar{k}_{i}}^{i}(x,y)\right)^{*}\left(d_{s\bar{k}_{s}}^{j}(x,y)\right)^{*}d_{p\bar{k}_{p}}^{l}(x,y)\\
 & =\frac{2}{\epsilon_{0}\hbar}\sqrt{\frac{\hbar^{3}\bar{\omega}_{i}\bar{\omega}_{s}\bar{\omega}_{p}}{(2)^{3}}}\int dxdy\ \frac{\chi_{2}^{ijl}(\mathbf{r})}{\epsilon_{0}^{2}n_{i}^{2}n_{j}^{2}n_{l}^{2}}\left(d_{i\bar{k}_{i}}^{i}(x,y)\right)^{*}\left(d_{s\bar{k}_{s}}^{j}(x,y)\right)^{*}d_{p\bar{k}_{p}}^{l}(x,y).
\end{align}
\end{subequations} Note the extra factor of two in the definition
of $\zeta_{s/i}(z)$ that comes about because of the permutation symmetry
of the $\Gamma$ coefficients. In the last equations we ignored the
$\chi_{2}$ contributions to $\Gamma_{3}$, but they can be easily
added. 

\section{Connecting the $\omega,t$ and $\omega,z$ operators}

\label{ders:app} We want to transform the equations of motion Eq.
(\ref{eomszt}), expressing them in the reciprocal frequency space
and in a frame of reference that propagates at the pump group velocity.
We begin by defining the $(t,\omega)$ Fourier transform of the field
operators 
\begin{align}
\tilde{\psi}_{\mu}(z,\omega) & =\sqrt{v_{\mu}}\int\frac{dt}{\sqrt{2\pi}}e^{i\omega t}\psi_{\mu}(z,t),\label{ftransform}\\
\psi_{\mu}(z,t) & =\int\frac{d\omega}{\sqrt{2\pi v_{\mu}}}e^{-i\omega t}~\tilde{\psi}_{\mu}(z,\omega).
\end{align}
Here we consider SPDC as the interaction generating twin-beams. Applying
$\int\frac{dt}{\sqrt{2\pi}}e^{i\omega t}$ to both sides of Eq. (\ref{eomszt})
and substituting the $z,t$ operators in terms of their Fourier transforms,
we find the equivalent equation  for the signal 
\begin{align}
\frac{\partial}{\partial z}\tilde{\psi}_{s}(z,\omega) & =i\left(\frac{\omega-\bar{\omega}_{s}}{v_{s}}\right)\tilde{\psi}_{s}(z,\omega)\label{fourier:dyn}\\
 & +i\int\frac{d\omega'}{\sqrt{2\pi v_{p}v_{s}v_{i}}}\xi_{1}(z)\ \langle\tilde{\psi}_{p}(z,\omega+\omega')\rangle\ \tilde{\psi}_{i}^{\dagger}(z,\omega')\\
 & +i\int\frac{d\omega'}{\sqrt{2\pi}v_{s}}\zeta_{s}(z)\ I_{0}(z,\omega-\omega')\ \psi_{s}(z,\omega'),
\end{align}
where we have defined the $(t,\omega)$ Fourier transform of the energy density of the pump in $z$ as
\begin{equation}
I_{0}(z,\omega)=v_{p}\int dt\ e^{i\omega t}\ |\langle\psi_{p}(z,t)\rangle|^{2}=e^{i\omega t_{0}}e^{i\omega z/v_{p}}\int dz'\ e^{-i\omega z'/v_{p}}\ |\Lambda(z)|^{2}.
\end{equation}
Now we make the following change of variables, moving to a frame of
reference that propagates at the pump group velocity, 
\begin{align}
 & \tilde{\psi}_{j}(z,\omega)=e^{i\frac{\omega-\bar{\omega}_{j}}{v_{p}}z}a_{j}(z,\omega)\ \ \ j\in\{s,i\},\label{chVarMovingFrame}\\
 & \langle\tilde{\psi}_{p}(z,\omega)\rangle=e^{i\frac{\omega-\bar{\omega}_{p}}{v_{p}}z}\beta_{p}(z,\omega).
\end{align}
We can see that the pump amplitude in this frame of reference is $z$-independent
(in the absence of SPM) by applying the solution to the pump dynamics
found in Eq. (\ref{pumpSol}). 
\begin{align}
\beta_{p}(z,\omega) & =\sqrt{\hbar\bar{\omega}_{p}}e^{-i\frac{\omega-\bar{\omega}_{p}}{v_{p}}z}\int\frac{dt}{\sqrt{2\pi v_{p}}}e^{i\omega t}\ \langle\psi_{p}(z,t)\rangle=\sqrt{\hbar\bar{\omega}_{p}}e^{-i\frac{\omega-\bar{\omega}_{p}}{v_{p}}z}\int\frac{dt}{\sqrt{2\pi v_{p}}}e^{i\omega t}\ \Lambda(z-v_{p}(t-t_{0}))\ e^{-i\bar{\omega}_{p}(t-t_{0})+i\varphi(z,t)},
\end{align}
which, when making the change of variables $z'=z-v_{p}t$, yields 
\begin{align}
\beta_{p}(z,\omega) & =\sqrt{\hbar\bar{\omega}_{p}}e^{i\bar{\omega}_{p}t_{0}}\int\frac{dz'}{\sqrt{2\pi/v_{p}}}e^{-i\frac{\omega-\bar{\omega}_{p}}{v_{p}}z'}\Lambda(z')\ e^{i\ \theta(z,z')},\label{pumpSolTransformed}
\end{align}
where $\theta(z,z')=\varphi(z,t_{0}+\frac{z-z'}{v_{p}})$. When SPM
is negligible, the nonlinear phase $\varphi(z,t)$ is negligible,
rendering Eq. (\ref{pumpSolTransformed}) independent of $z$.

\
 The change of variables in Eq. (\ref{chVarMovingFrame}) yields the
following equation  for the signal

\begin{align}
\frac{\partial}{\partial z}a_{s}(z,\omega) & =i(\omega-\bar{\omega}_{s})\left(\frac{1}{v_{s}}-\frac{1}{v_{p}}\right)a_{s}(z,\omega)\label{der1}\\
 & +i\int\frac{d\omega'}{\sqrt{2\pi\hbar\bar{\omega_{p}}v_{s}v_{i}v_{p}}}\xi_{1}(z)e^{i\frac{\bar{\omega}_{s}+\bar{\omega}_{i}-\bar{\omega}_{p}}{v_{p}}z}\beta_{p}(z,\omega+\omega')a_{i}^{\dagger}(z,\omega')\\
 & +i\ \int\frac{d\omega'}{2\pi\hbar\bar{\omega}_{p}v_{s}v_{p}}\zeta_{s}(z)\mathcal{E}_p\omega-\omega')a_{s}(z,\omega'),
\end{align}
where we have defined 
\begin{equation}
\mathcal{E}_p(\omega)=\hbar\bar{\omega}_{p}I_{0}(z,\omega)e^{-i\omega z/v_{p}}=\hbar\bar{\omega}_{p}e^{i\omega t_{0}}\int\frac{dz'}{\sqrt{2\pi}}\ e^{-i\omega z'}\ |\Lambda(z)|^{2},
\end{equation}
which is \emph{always} independent of $z$, regardless of the SPM
of the pump. Note that we can further simplify Eq. (\ref{der1}) 
by noting that $\bar{\omega}_{s}+\bar{\omega}_{i}-\bar{\omega}_{p}=0$.

Finally, let us consider the case where the process is phase-matched
for SFWM. In this case we define Fourier transformed operators for
the signal and idler fields as in Eq. (\ref{ftransform}). However,
for the pump we define 
\begin{align}
\phi_{p}(z,\omega) & =\sqrt{v_{p}}\int\frac{dt}{\sqrt{2\pi}}e^{i\omega t}\braket{\psi_{p}(z,t)}^{2},\quad\braket{\psi_{p}(z,t)}^{2}=\int\frac{d\omega}{\sqrt{2\pi v_{p}}}e^{-i\omega t}~\phi_{p}(z,\omega).
\end{align}
In terms of $\phi$, the new equation of motion for $\tilde{\psi}_{s}$
has the same form as Eq. (\ref{fourier:dyn}) with the replacement
$\xi_{1}(z)\braket{\tilde{\psi}_{p}(z,\omega+\omega')}\to\xi_{2}(z)\phi_{p}(z,\omega+\omega')$.
We can shift to a frame moving at the pump group velocity as we did
in Eq. (\ref{chVarMovingFrame}), but for the pump we define
\begin{align}
\beta_{p}(z,\omega) & =\hbar\bar{\omega}_{p}e^{-i\frac{\omega-2\bar{\omega}_{p}}{v_{p}}z}\int\frac{dt}{\sqrt{2\pi/v_{p}}}e^{i\omega t}\ \langle\psi_{p}(z,t)\rangle^{2}=e^{i\omega t_{0}}\frac{(\hbar\bar{\omega}_{p})}{\sqrt{2\pi v_{p}}}\int dz'e^{-iz'\frac{(\omega-2\bar{\omega}_{p})}{v_{p}}}\left(\Lambda(z')\right)^{2}e^{i2\theta(z,z'))}.
\end{align}
Note the factor of two multiplying $\bar{\omega}_{p}$ and exponentiating
$\braket{\psi_{p}(z,t)}$. With these definitions we arrive at an
equation  analogous to Eq. (\ref{der1}), but where we need to replace
\begin{align}
\xi_{1}(z)\frac{1}{\sqrt{\hbar\bar{\omega}_{p}}}e^{i\frac{\bar{\omega}_{s}-\bar{\omega}_{i}-\bar{\omega}_{p}}{v_{p}}z}\to\frac{1}{\hbar\bar{\omega}_{p}}\xi_{2}(z)e^{i\frac{\bar{\omega}_{s}-\bar{\omega}_{i}-2\bar{\omega}_{p}}{v_{p}}z}.
\end{align}
However,  for SFWM one has $\bar{\omega}_{s}-\bar{\omega}_{i}-2\bar{\omega}_{p}=0$. 

\section{Connecting free operators in space and time}

\label{inout:app} We use the following definitions \begin{subequations}
\begin{align}
\psi_{j}(z,t) & =\int\frac{d\omega}{\sqrt{2\pi v_{j}}\sqrt{v_{j}}}e^{i(\omega-\bar{\omega}_{j})z/v_{j}}b_{jk_{j}(\omega)}(t),\quad k_{j}(\omega)\equiv\bar{k}_{j}+(\omega-\bar{\omega}_{j})\label{def1}\\
 & =\int\frac{d\omega}{\sqrt{2\pi v_{j}}}e^{i(\omega-\bar{\omega}_{j})z/v_{p}}e^{-i\omega t}c_{j}(z,\omega)\label{def2}
\end{align}
\end{subequations} It is useful to label spacetime coordinates $(t_{0},z_{0})$
as ``distant past'' if $t_{0}$ is a time before the nonlinear interaction
has effect and $z_{0}$ is a coordinate less than coordinates where
the nonlinearity is present, and to label spacetime coordinates $(t_{1},z_{1})$
as ``distant future'' if $t_{1}$ is a time after the nonlinear interaction
has effect and $z_{1}$ is a coordinate greater than coordinates where
the nonlinearity is present. In Appendix \ref{solzt} we show that
for $(t_{n},z_{n})$ either in the distant past or distant future
we have 
\begin{align}
\psi_{j}(z_{n},t)=e^{-i\bar{\omega}_{j}(t-t_{n})}\psi_{j}(z_{n}-v_{j}(t-t_{n}),t_{n}).
\end{align}
Now we can use Eq. (\ref{def2}) for the LHS of the last equation
and Eq. (\ref{def1}) for the RHS to find 
\begin{align}
\psi_{j}(z_{n},t) & =e^{-i\bar{\omega}_{j}(t-t_{n})}\psi(z_{n}-v_{j}(t-t_{n}),t_{n})\\
\int\frac{d\omega}{\sqrt{2\pi v_{j}}}e^{i(\omega-\bar{\omega}_{j})z_{n}/v_{p}}e^{-i\omega t}c_{j}(z_{n},\omega) & =e^{-i\bar{\omega}_{j}(t-t_{n})}\int\frac{d\omega}{\sqrt{2\pi v_{j}}\sqrt{v_{j}}}b_{jk_{j}(\omega)}(t_{n})e^{i(\omega-\bar{\omega}_{j})(z_{n}-v(t-t_{n}))/v_{j}}\\
 & =\int\frac{d\omega}{\sqrt{2\pi v_{j}}\sqrt{v_{j}}}e^{i(\omega-\bar{\omega}_{j})z_{n}/v_{j}}e^{-i\omega t}e^{i\omega t_{n}}b_{jk_{j}(\omega)}(t_{0}).
\end{align}
Comparing the quantities under the integral we see  that 
\begin{align}
a_{j}(z_{n},\omega) e^{-i\Delta k_{j}(\omega)z_{n}}=e^{i\omega t_{n}}b_{jk_{j}(\omega)}(t_{n})/\sqrt{v_{j}}.
\end{align}

\section{Formal solution in $(z,t)$}

\label{solzt} We will construct an implicit solution of the $(t,z)$
equations of motion, where we introduce spacetime points $(t_{n},z_{n})$
in the distant past $(n=0$) or the distant future $(n=1)$, where
these terms are defined in the Appendix above.  For $n=0$ or $n=1$
we can write a formal solution of the equations In either case we
one can write a formal solution of the propagation equation (\ref{psi1Heisenberg}),
\begin{align}
\bar{\psi}_{s}(z,t) & =\bar{\psi}_{s}(z-v_{s}(t-t_{n}),t_{n})\label{fsol}\\
 & +\frac{\theta(t-t_{n})}{v_{s}}\int_{z_{-}}^{z}dz'f(z',t-\tfrac{z-z'}{v_{s}})\bar{\psi}_{i}^{\dagger}(z',t-\tfrac{z-z'}{v_{s}})\nonumber \\
 & +\frac{\theta(t-t_{n})}{v_{s}}\int_{z_{-}}^{z}dz'g(z',t-\tfrac{z-z'}{v_{s}})\bar{\psi}_{s}(z',t-\tfrac{z-z'}{v_{s}})\nonumber \\
 & -\frac{\theta(t_{n}-t)}{v_{s}}\int_{z}^{z_{-}}dz'f(z',t-\tfrac{z-z'}{v_{s}})\bar{\psi}_{i}^{\dagger}(z',t-\tfrac{z-z'}{v_{s}})\nonumber \\
 & -\frac{\theta(t_{n}-t)}{v_{s}}\int_{z}^{z_{-}}dz'g(z',t-\tfrac{z-z'}{v_{s}})\bar{\psi}_{s}(z',t-\tfrac{z-z'}{v_{s}})\nonumber 
\end{align}
where we defined \begin{subequations} 
\begin{align}
\bar{\psi}_{j}(z,t) & =e^{i\bar{\omega}_{j}t}\psi_{j}(z,t)\\
z_{-} & =z-v_{s}(t-t_{n}),\\
f(z,t) & =\xi_{\delta}(z)\braket{\psi_{p}(z,t)}^{\delta},\\
g(z,t) & =\zeta_{s}(z)|\braket{\bar{\psi}_{p}(z,t)}|^{2},
\end{align}
\end{subequations} 
and $\theta(t)$ is the Heaviside step function, $\theta(t) = 0$ if $t<0$, $\theta(t) = 1$ if $t>0$ and $\theta(t) = 1/2$ if $t=0$.

First, we investigate the distant past case (setting
$n=0$). We can introduce a new dummy integration variables for the
integrals extending from $z$ to $z_{-}$ as follows 
\begin{align}
t'=t-\frac{z-z'}{v_{s}}\label{dummy}\\
z'=z+v_{s}(t'-t)
\end{align}
and when $z'=z$ we have $t'=t$ and when $z'=z_{-}$ we have $t'=t_{0}$.
With this change of variables we have 
\begin{align}
 & \bar{\psi}_{s}(z_{0},t)=\bar{\psi}_{s}(z_{0}-v_{s}(t-t_{0}),t_{0})\label{eq:AGreen_mod}\\
 & +\frac{\theta(t-t_{0})}{v_{s}}\int_{z_{-}}^{z_{0}}f(z',t-\tfrac{z_{0}-z'}{v_{s}})\bar{\psi}_{i}^{\dagger}(z',t-\tfrac{z_{0}-z'}{v_{s}})dz'\nonumber \\
 & +\frac{\theta(t-t_{0})}{v_{s}}\int_{z_{-}}^{z_{0}}g(z',t-\tfrac{z_{0}-z'}{v_{s}})\bar{\psi}_{s}(z',t-\tfrac{z_{0}-z'}{v_{s}})dz'\nonumber \\
 & -\theta(t_{0}-t)\int_{t}^{t_{0}}f(z_{0}+v_{s}(t'-t),t')\bar{\psi}_{i}^{\dagger}(z+v_{s}(t'-t),t')dt'\nonumber \\
 & -\theta(t_{0}-t)\int_{t}^{t_{0}}g(z_{0}+v_{s}(t'-t),t')\bar{\psi}_{s}(z+v_{s}(t'-t),t')dt'.\nonumber 
\end{align}
For $(t_{0},z_{0})$ in the distant past  the spatial extent of the
classical pump $\bar{\psi}_{p}(z,t_{0})$ has zero overlap with the
nonlinear region, and $z_{0}$ is smaller than the values of the arguments
for which  the nonlinear coefficients $\zeta_{j}(z)$, $\xi(z)$ are
nonzero, so we  have 
\begin{align}
\bar{\psi}_{j}(z_{0},t) & =\bar{\psi}_{j}(z_{0}-v_{j}(t-t_{0}),t_{0}),\label{z0}\\
\psi_{j}(z_{0},t) & =e^{-i\bar{\omega}_{j}(t-t_{0})}\psi_{j}(z_{0}-v_{j}(t-t_{0}),t_{0})
\end{align}
This is readily established by noticing that the first pair of integrals
on the right-hand side of Eq. (\ref{eq:AGreen_mod}) range over values
of $z'$ for which $f(z',t-\tfrac{z_{o}-z'}{v_{s}})$ and $g_{s}(z',t-\tfrac{z_{o}-z'}{v_{s}})$
will vanish (seen by examining the range of the first argument since
$z'<z_{0}$), and the last pair of integrals on the right-hand-side
will range over values of $t'$ for which $f(z_{0}+v_{s}(t'-t),t')$
and $g(z_{0}+v_{s}(t'-t),t')$ will vanish (seen by examining the
range of the second argument since now $t'<t_{0}$).

Now let us study the distant future solution $(n=1$). The formal
solution corresponding to (\ref{eq:AGreen_mod}) is then 
\begin{align}
 & \bar{\psi}_{s}(z_{1},t)=\bar{\psi}_{s}(z_{1}-v_{s}(t-t_{1}),t_{1})\\
 & +\theta(t-t_{1})\int_{t_{1}}^{t}dt'f(z_{1}+v_{s}(t'-t),t')\bar{\psi}_{i}^{\dagger}(z_{1}+v_{s}(t'-t),t')\nonumber \\
 & +\theta(t-t_{1})\int_{t_{1}}^{t}dt'g(z_{1}+v_{s}(t'-t),t')\bar{\psi}_{s}(z_{1}+v_{s}(t'-t),t')\nonumber \\
 & -\frac{\theta(t_{1}-t)}{v_{s}}\int_{z_{1}}^{z_{-}}dz'f(z',t-\tfrac{z_{1}-z'}{v_{s}})\bar{\psi}_{i}^{\dagger}(z',t-\tfrac{z_{1}-z'}{v_{s}})\nonumber \\
 & -\frac{\theta(t_{1}-t)}{v_{s}}\int_{z_{1}}^{z_{-}}dz'g(z',t-\tfrac{z_{1}-z'}{v_{s}})\bar{\psi}_{s}(z',t-\tfrac{z_{1}-z'}{v_{s}}).\nonumber 
\end{align}
Using arguments similar to those just made for distant past times $(n = 0)$, we arrive at the
corresponding results for distant future times $(n = 1)$:

\begin{align}
\bar{\psi}_{j}(z_{1},t) & =\bar{\psi}_{j}(z_{1}-v_{j}(t-t_{1}),t_{1})\label{z1},\\
\psi_{j}(z_{1},t) & =e^{-i\bar{\omega}_{j}(t-t_{1})}\psi_{j}(z_{1}-v_{j}(t-t_{1}),t_{1}).
\end{align}

\section{Low gain solutions}

\label{lowgain} We go back to Eq. (\ref{eomszw}) and solve this
equations perturbatively for the case of SPDC (and assuming no XPM).
First we define the operators 
\begin{align}
c_{j}(z,\omega)=e^{i\Delta k_j(\omega)z}a_{s}(z,\omega).
\end{align}
Using these definitions in  Eq. (\ref{eomszw}) and integrating to
first order we find 
\begin{align}
c_{s}(z_{1},\omega) & =c_{s}(z_{0},\omega)+i\int_{z_{0}}^{z_{1}}dz\int d\omega'\beta_{p}(z,\omega+\omega')e^{-iz(\Delta k_{s}(\omega)+\Delta k_{i}(\omega'))}c_{i}^{\dagger}(z,\omega'),\label{transp}\\
c_{i}^{\dagger}(z_{1},\omega) & =c_{i}^{\dagger}(z_{0},\omega)-i\int_{z_{0}}^{z_{1}}dz\int d\omega'\beta_{p}^{*}(z,\omega+\omega')e^{iz(\Delta k_{s}(\omega)+\Delta k_{i}(\omega'))}c_{s}^{\dagger}(z,\omega').
\end{align}
Now we assume the nonlinear interaction is weak and replace $c_{i}(z,\omega')\approx c(z_{0},\omega')$
in the RHS. Furthermore we assume that the pump spectral amplitude
is not modified by SPM, and thus that there is no $z$ dependence
in $\beta_{p}$. We introduce the net phase mismatch 
\begin{align}
\Delta k(\omega,\omega')=\Delta k_{s}(\omega)+\Delta k_{i}(\omega)=\frac{\omega-\bar{\omega}_{s}}{v_{s}}+\frac{\omega'-\bar{\omega}_{i}}{v_{i}}-\frac{\omega+\omega'-\bar{\omega}_{p}}{v_{p}},
\end{align}
and we can then write the transformation in Eq. (\ref{transp}) as
\begin{align}
c_{s}(z_{1},\omega) & =\int\bar{U}^{s,s}(\omega,\omega';z_{1},z_{0})c_{s}(z_{0},\omega')+\int\bar{U}^{s,i}(\omega,\omega';z_{1},z_{0})c_{i}^{\dagger}(z_{0},\omega'),\\
c_{i}^{\dagger}(z_{1},\omega) & =\int(\bar{U}^{i,s}(\omega,\omega';z_{1},z_{0}))^{*}c_{s}(z_{0},\omega')+\int(\bar{U}^{i,i}(\omega,\omega';z_{1},z_{0}))^{*}c_{i}^{\dagger}(z_{0},\omega'),
\end{align}
where the perturbative transfer functions can be jointly decomposed
as follows: 
\begin{align}
\bar{U}^{s,s}(\omega,\omega';z_{1},z_{0}) & =\sum_{k}\cosh(r_{k})[\rho_{s}^{(k)}(\omega)][\rho_{s}^{(k)}(\omega')]^{*},\\
\bar{U}^{s,i}(\omega,\omega';z_{1},z_{0}) & =\sum_{k}\sinh(r_{k})[\rho_{s}^{(k)}(\omega)][\rho_{i}^{(k)}(\omega')]^{*},\\
(\bar{U}^{i,i}(\omega,\omega';z_{1},z_{0}))^{*} & =\sum_{k}\cosh(r_{k})[\rho_{i}^{(k)}(\omega)]^{*}[\rho_{i}^{(k)}(\omega')],\\
(\bar{U}^{i,s}(\omega,\omega';z_{1},z_{0}))^{*} & =\sum_{k}\sinh(r_{k})[\rho_{i}^{(k)}(\omega)]^{*}[\rho_{s}^{(k)}(\omega')]^{*}.
\end{align}
Here the functions $\rho_{s/i}$ are complete and orthonormal, and
are the Schmidt functions of the joint spectral amplitude 
\begin{align}
J(\omega,\omega')=\frac{1}{\sqrt{ v_{s}v_{i}v_{p}}}\beta_{p}(\omega+\omega')\Phi(\Delta k (\omega,\omega'))=\sum_{l}r_{l}\rho_{s}^{(l)}(\omega)\rho_{s}^{(l)}(\omega),\\
\Phi(\Delta k(\omega,\omega'))=\int_{z_{0}}^{z_{1}}\frac{dz}{\sqrt{2\pi}}e^{-iz\Delta k(\omega,\omega')}\xi_{1}(z),
\end{align}
in the approximation that the squeezing parameters $r_{l}\ll1$ and
thus $\sinh(r_{l})\approx r_{l}$ and $\cosh(r_{l})\approx1$. Note
that in this limit we  recover the well known result that the JSA
is simply the product of the pump and phase matching function.

Comparing  the results of this appendix with the more general expression
in Eq. (\ref{expansion1}) obtained for arbitrary gain, we see  that
in the low-gain regime $\tau_{s/i}(\omega)=\rho_{s/i}(\omega)$

\section{The problem with group velocity dispersion}

\label{app:disp}

\label{app:comms} We investigate how group velocity dispersion
modifies the conclusions drawn in this paper. In particular we will
consider how equal position and different time commutators such
as 
\begin{align}
[\bar{\psi}_{i}^{\dagger}(z,t),\bar{\psi}_{s}^{\dagger}(z,t_{0})]\label{expansion}
\end{align}
are modified by the inclusion of group velocity dispersion. For the
sake of concreteness we will assume that one is only interested in
SPDC, and that XPM can be assumed to be  unimportant. We can then
write the generalized form of the equations of motion (\ref{psi1Heisenberg},\ref{psi2Heisenberg})
for the field operators as 
\begin{align}
\left(\frac{\partial}{\partial t}+v_{s}\frac{\partial}{\partial z}+i\frac{v_{s}'}{2}\frac{\partial^{2}}{\partial z^{2}}\right)\bar{\psi}_{s}(z,t) & =if(z,t)\bar{\psi}_{i}^{\dagger}(z,t),\label{eq:dynamics_use-2}\\
\left(\frac{\partial}{\partial t}+v_{i}\frac{\partial}{\partial z}+i\frac{v_{i}'}{2}\frac{\partial^{2}}{\partial z^{2}}\right)\bar{\psi}_{i}^{\dagger}(z,t) & =-if^{*}(z,t)\bar{\psi}_{s}(z,t),\nonumber 
\end{align}
where $f(z,t) = \xi_1 (z) \braket{\bar{\psi}_p(z,t)}$, and we have assumed
that the group velocity dispersion $v_{s}'$ and $v_{i}'$ (see Eqs.
(\ref{eq:GV=000026GVD})) can be taken as independent of $k$. In
the case of no nonlinearity one can write the formal solution of this
problem in terms of a Green function 
\begin{align}
\bar{\psi}_{j}(z,t)=\int dz'G_{j}(z-z';t-t-t_{0})\bar{\psi}_{j}(z,t_{0}),
\end{align}
where 
\begin{align}
G_{j}(z;t)=\frac{(1-i\text{sign}(v'_{j}t))}{\sqrt{4\pi\left|v_{j}'t\right|}}e^{\left(\frac{i(z-v_{j}t)^{2}}{2v'_{j}t}\right)}.\label{eq:Green_GVD}
\end{align}
Note that in the limit $v'_{j}\to0$ the last equation collapses to
\begin{align}
G_{j}(z;t)=\delta(z-v_{j}t).
\end{align}
Using the Green functions we can write a formal solution of the equations
of motion including the nonlinearity as follows 
\begin{align}
 & \bar{\psi}_{s}(z,t)=\int G_{s}(z-z';t-t_{o})\bar{\psi}_{s}(z',t_{o})dz'\label{eq:AGreen_use}\\
 & +i\int G_{s}(z-z';t-t')\Theta(t-t',t_{o}-t')f(z',t')\bar{\psi}_{i}^{\dagger}(z',t')dz'dt'\nonumber 
\end{align}
where 
\begin{align}
\Theta(t_{2};t_{1})\equiv\theta(t_{2})-\theta(t_{1}),
\end{align}
and a similar equation for $\bar{\psi}_{i}^{\dagger}(z,t)$.

Having constructed an implicit solution we can develop a perturbation
theory in which on the right hand side of the last equation we iteratively
replace the ``evolved'' time fields $\bar{\psi}_{j}(z,t),t\neq t_{0}$
under the integral. To first order in the nonlinearity we find 
\begin{align}
 & \bar{\psi}_{s}(z,t)=\int G_{s}(z-z';t-t_{o})\bar{\psi}_{s}(z',t_{o})dz'\\
 & +i\int G_{s}(z-z';t-t')\Theta(t-t';t'-t_{o})f(z',t')G_{i}^{*}(z'-z'';t'-t_{o})\bar{\psi}_{i}^{\dagger}(z'',t_{o})dz'dz''dt'+....
\end{align}
Using the expansion for the fields we find that the commutator in
Eq. (\ref{expansion}) is 
\begin{align}
 & \left[\bar{\psi}_{i}^{\dagger}(z,t),\bar{\psi}_{s}^{\dagger}(z_{o},t_{o})\right]\approx-i\int F(z,z_{o};t,t_{o},t')\Theta(t-t';t'-t_{o})dt',\label{eq:approxB}\\
 & F(z,z_{o};t,t_{o},t')=\int G_{i}^{*}(z-z';t-t')f^{*}(z',t')G_{s}(z'-z_{o};t'-t_{o})dz'.\label{eq:Fdef}
\end{align}
In the next sections we evaluate this quantity in two limits.

\subsection{No group velocity dispersion}

Using the results for the case of no GVD we find 
\begin{align*}
 & F(z,z_{o};t,t_{o},t')=\int\delta(z-z'-v_{i}(t-t'))g^{*}(z',t')\delta(z'-z_{o}-v_{s}(t'-t_{o})dz'\\
 & =\delta(z-z_{o}-v_{s}(t'-t_{o})-v_{i}(t-t'))g^{*}(z_{o}+v_{s}(t'-t_{o})).
\end{align*}
Of particular interest for our Fourier transform variables is the
equal position commutator 
\begin{align}
 & \left[\bar{\psi}_{i}^{\dagger}(z,t),\bar{\psi}_{s}^{\dagger}(z,t_{o})\right]\approx\int F(z,z;t,t_{o},t')\Theta(t-t';t'-t_{o})dt'\label{eq:noGVDlimit}\\
 & =\int\delta(-v_{s}(t'-t_{o})-v_{i}(t-t'))g^{*}(z+v_{s}(t'-t_{o}))\Theta(t-t';t'-t_{o})dt'.\nonumber 
\end{align}
But since $v_{s}$ and $v_{i}$ are both positive the Dirac delta
function will only give a contribution at values of $t'$ where the
$\Theta$ function vanishes, and so we have 
\begin{align*}
 & \left[\bar{\psi}_{i}^{\dagger}(z,t),\bar{\psi}_{s}^{\dagger}(z,t_{o})\right]\approx0,
\end{align*}
and so the equal position commutators in the presence of the pump
are, at least to first order, equivalent to the equal position commutators
in the absence of the pump. We saw in the text that,  for no group
velocity dispersion, this equivalence holds to all orders in the presence
of the pump.

\subsection{Finite group velocity dispersion}

In this case after some lengthy algebra one finds 
\begin{align}
 & F(z,z;t,t_{o},t')=\label{eq:Fwork4}\\
 & \frac{e^{-\frac{2i\mathcal{A}^{2}}{D}}}{2\pi\sqrt{\left|\tau_{1}\tau_{2}v'_{s}v'_{i}\right|}}\left(\frac{(1+i\text{sign}(D))}{\sqrt{2}}\right)\left(\frac{(1-i\text{sign}(\frac{\tau_{1}\tau_{2}v'_{s}v'_{i}}{D}))}{\sqrt{2}}\right)\int e^{\frac{iD\left(\mathcal{D}-\frac{4\mathcal{A}A}{D}-2(z-z')\right)^{2}}{8\tau_{1}\tau_{2}v'_{s}v'_{i}}}f^{*}(z',t')dz'.\nonumber 
\end{align}
where 
\begin{align}
 & \tau_{1}=t-t',\\
 & \tau_{2}=t'-t_{0},\\
 & \mathcal{D}=\tau_{1}v_{i}-\tau_{2}v_{s},\\
 & \mathcal{A}=\frac{1}{2}\left(\tau_{1}v_{i}+\tau_{2}v_{s}\right),\\
 & D=\tau_{1}v'_{i}-\tau_{2}v'_{s},\\
 & A=\frac{1}{2}\left(\tau_{1}v'_{i}+\tau_{2}v'_{s}\right),
\end{align}
The last integral can be  evaluated asymptotically in the limit that
the $v_{j}'$ are ``small,'' (cf. Sec. 2.9 of Erd\'elyi \cite{erdelyi1956asymptotic}) and one finds 
\begin{align*}
 & F(z,z;t,t_{o},t')\sim\left(\frac{(1+i \text{sign}(D))}{\sqrt{2}}\right)\frac{1}{\sqrt{2\pi\left|D\right|}}e^{-\frac{2i\mathcal{A}^{2}}{D}}g^{*}(\bar{z},t'),
\end{align*}
and then 
\begin{align}
 & \left[\bar{\psi}_{i}^{\dagger}(z,t),\bar{\psi}_{s}^{\dagger}(z,t_{o})\right]\sim i\int\left(\frac{(1+i\text{sign}(D))}{\sqrt{2}}\right)\frac{1}{\sqrt{2\pi\left|D\right|}}e^{-\frac{2i\mathcal{A}^{2}}{D}}f^{*}(\bar{z},t')\Theta(t-t';t'-t_{o})dt'.\label{eq:comm_asympt}
\end{align}
Again one can take the limit of vanishing group velocity dispersion by putting $\left|D\right|\rightarrow0$, in which limit  
\begin{align*}
 & \sqrt{\frac{2}{\pi\left|D\right|}}\left(\frac{1+i\text{sign}(D)}{\sqrt{2}}\right)e^{-\frac{2i\mathcal{A}^{2}}{D}}\rightarrow\delta(\mathcal{A}),
\end{align*}
and using this in (\ref{eq:comm_asympt}) we have have 
\begin{align*}
 & \left[\bar{\psi}_{i}^{\dagger}(z,t),\bar{\psi}_{s}^{\dagger}(z,t_{o})\right]\rightarrow\frac{1}{2}\int\delta(\mathcal{A})f^{*}(z-\tfrac{\mathcal{D}}{2},t')\Theta(t-t';t'-t_{o})dt'\\
 & =\frac{1}{2}\int\delta\left(\frac{1}{2}\left(\tau_{1}v_{i}+\tau_{2}v_{s}\right)\right)f^{*}(z-\tfrac{\tau_{1}v_{i}-\tau_{2}v_{s}}{2})\Theta(t-t';t'-t_{o})dt'\\
 & =\int\delta\left(\tau_{1}v_{i}+\tau_{2}v_{s}\right)f^{*}(z-\tfrac{\tau_{1}v_{i}-\tau_{2}v_{s}}{2})\Theta(t-t';t'-t_{o})dt'\\
 & =\int\delta\left(\tau_{1}v_{i}+\tau_{2}v_{s}\right)f^{*}(z+v_{s}\tau_{2})\Theta(t-t';t'-t_{o})dt'\\
 & =\int\delta(v_{i}(t-t')+v_{s}(t'-t_{o})f^{*}(z+v_{s}(t'-t_{o}))\Theta(t-t';t'-t_{o})dt' =0,
\end{align*}
indeed in agreement with the limit (\ref{eq:noGVDlimit}) of vanishing
group velocity dispersion, as expected. But for a finite group velocity
dispersion (\ref{eq:comm_asympt}) indicates that we cannot expect
this commutator to vanish.  
\end{document}